\documentclass[11pt]{article}

\usepackage{fullpage}

\usepackage{url}
\usepackage{booktabs} % For formal tables
\usepackage{multirow} % For merged cell in tables
\usepackage{multicol}
\usepackage{diagbox} % For backslash column in tables
\usepackage{array}
\usepackage{comment}
\newcolumntype{x}[1]{>{\centering\arraybackslash}p{#1}}
\newcolumntype{y}[1]{>{\arraybackslash}p{#1}}

\usepackage{listings}

\usepackage{graphicx}
\usepackage{caption}
\usepackage{subcaption}

\usepackage{color}
\definecolor{green}{rgb}{0, 0.42, 0.24}
\definecolor{blue}{rgb}{0, 0.53, 0.74}

\usepackage{algorithm}% http://ctan.org/pkg/algorithms
\usepackage{algpseudocode}% http://ctan.org/pkg/algorithmicx

\usepackage{amsmath}
\DeclareMathOperator*{\argmax}{arg\,max}
\DeclareMathOperator*{\argmin}{arg\,min}

\begin{document}
\title{\vspace*{-0.8cm} \textsc{PhishSim}: Aiding Phishing Website Detection \\ with a Feature-Free Tool}

\author{Rizka Purwanto$^{1,3,4}$, Arindam Pal$^{1,2,3,4}$, Alan Blair$^{1,4}$, Sanjay Jha$^{1,4}$ \\[0.2cm]
  $^1\,$School of Computer Science and Engineering, University of New South Wales, Sydney 2052 \\
  $^2\,$CSIRO Data61, Sydney, Australia
  $^3\,$Cyber Security Cooperative Research Centre, Australia \\
  $^4\,$UNSW Institute for Cyber Security ({\large IFCYBER})
}

\markboth{IEEE Transactions on Information Forensics and Security,~Vol.~00, No.~0, August~2021}%
{Shell \MakeLowercase{\textit{et al.}}: Bare Demo of IEEEtran.cls for IEEE Journals}



\maketitle

\begin{abstract}
  In this paper, we propose a feature-free method for detecting phishing websites using the Normalized Compression Distance (NCD), a parameter-free similarity measure which computes the similarity of two websites by compressing them, thus eliminating the need to perform any feature extraction. It also removes any dependence on a specific set of website features. This method examines the HTML of webpages and computes their similarity with known phishing websites, in order to classify them.{\let\thefootnote\relax\footnote{Published in IEEE Transactions on Information Forensics \& Security 17, 1497-1512, 2022.}}

We use the Furthest Point First algorithm to perform phishing prototype extractions, in order to select instances that are representative of a cluster of phishing webpages. We also introduce the use of an incremental learning algorithm as a framework for continuous and adaptive detection without extracting new features when concept drift occurs. On a large dataset, our proposed method significantly outperforms previous methods in detecting phishing websites, with an AUC score of 98.68\%, a high true positive rate (TPR) of around 90\%, while maintaining a low false positive rate (FPR) of 0.58\%. Our approach uses prototypes, eliminating the need to retain long term data in the future, and is feasible to deploy in real systems with a processing time of roughly 0.3 seconds.
\end{abstract}

%\begin{IEEEkeywords}
%phishing detection, webpage, incremental learning, feature-free methods.
%\end{IEEEkeywords}

% \fi
%\IEEEpeerreviewmaketitle

\section{Introduction}
%\IEEEPARstart
Phishing is defined as a cyber-attack which uses social engineering via digital means to persuade victims to disclose their personal information, such as their password or credit card number \cite{apwg2016}. \color{black}The strategies used in phishing attacks exploit human vulnerabilities in distinguishing between authentic and phishing messages or websites \cite{neupane2016neural}. \color{black} Phishing is a low-cost, yet essential tool to aid various cyber-attacks, as it is often used as the key step in advanced persistent threats. With our ever-increasing reliance on various digital platforms, phishing has become a versatile weapon in the attacker's arsenal. Despite the general definition of phishing, the term itself has been commonly associated specifically with Web phishing attacks that use emails or SMS as the attack vector to lure victims into submitting personal information via phishing websites or by downloading malicious software. These websites are typically crafted to look professional and convincing as if they are legitimate.

As mentioned in their report, the Anti-Phishing Working Group (APWG) recorded a considerable increase in unique phishing attacks from 2014 to 2016 \cite{apwg2016}, which caused significant financial losses, estimated to be between \$60 million and \$3 billion per year in the United States \cite{Hong2012}. In another report, the APWG detected around 65,400 phishing websites per month in 2018 \cite{apwg2018}, while PhishLabs reported that phishing volume grew by 40.9\% in 2018 compared to the previous years \cite{phishlabs2019}. PhishLabs also reported how attack volumes continue to increase as actors' methods evolve and adapt to changes in the digital landscape. Furthermore, the use of free hosting providers has resulted in an increase in phishing attacks during the past four years, from 3.0\% in 2015 to 13.8\% in 2018. It is also relatively easy to set up phishing websites using phishing toolkits. The availability of these toolkits enables a single actor to create a large number of professional-looking phishing websites in a short period of time. As an example, PhishLabs reported a sharp increase in the number of attacks in August 2018. This phishing campaign used at least 2,000 freely-hosted phishing websites, all created using the same kit \cite{phishlabs2019}. With the availability of these phishing kits provided by organized crime groups, it is likely that the number of phishing attacks will increase further in the future. The use of free hosts, phishing kits, and SSL certificates show that there has been a persistent effort by actors to capitalize upon new opportunities, resulting in the continuous growth of the number of phishing attacks from year to year. This makes it challenging to develop a reliable phishing detection method that can deal with the dynamic nature of the attacks \cite{Ma:2009:ISU:1553374.1553462}.

To prevent the negative impacts of phishing, researchers have studied various methods in recent years to build an automated phishing website detection system by investigating the website content, appearance, URL, and other web-related features \cite{fu2006detecting, Whittaker2010, Xiang2011, xiang2009hybrid, zhang2011textual, Zhang2007cantina}. In general, these approaches could be classified into two categories. The first approach finds intrinsic features of phishing websites and attempts to detect these attacks based on these specific characteristics. Many studies in recent years go along with this type of approach, using machine learning and deep learning methods. While these methods have shown to perform well in detecting phishing, their detection are less robust towards concept drift since they are heavily based on certain characteristics assumed to be related to phishing websites (e.g., specific types of web forms or unusual structures in the URLs) which may change and become irrelevant in the future. Meanwhile, the second approach tries to detect phishing by measuring similarities between phishing websites and the targeted legitimate website. This method is less robust to zero-day phishing attacks compared to the first approach. However, the similarity-based methods are effective for quickly filtering a large amount of phishing websites, before being fed into machine learning based methods which typically take more time to perform classification. A number of previous studies have proposed the use of various similarity metrics and models to detect near similar phishing websites. These similarity-based techniques usually require the websites to be modeled into a certain representation space, e.g.\ using DOM trees, bag-of-words, or Doc2Vec models \cite{feng2020detection}.

In contrast, we attempt to propose a feature-free method for detecting phishing websites which uses the Normalized Compression Distance \cite{cilibrasi2005clustering} to compute website HTML similarity. The rationale of the study is based on the work of Cui et al. \cite{cui2017tracking}, which reveals the commonalities between phishing websites, demonstrating that 90\% of the 19,066 confirmed phishing websites from PhishTank \cite{phishTank} are replicas or variations of other already known phishing websites. In their study, Cui et al. proposes a new distance metric, the proportional distance, to assess website similarities for detecting phishing. This distance metric takes into account the number of occurrences of a predefined set of HTML tags, which are found to provide significant information whether a website is related to phishing attacks. On the other hand, the use of prototypes and the Normalized Compression Distance in our proposed method would eliminate the need of the predefined HTML tags, and uses compression algorithms to universally measure similarities between two streams of data based on the amount of information they have in common. \color{black}Our proposed method is not limited to a specific kind of phishing attack or phishing campaign, nor a certain type of the phishing email. However, our method is limited to variations of phishing websites which have occurred at least once, and unable to identify a new phishing website with a completely different and distinctive HTML structure.\color{black}

To summarize, this paper makes the following contributions:
\begin{itemize}
    \item We introduce a systematic method to perform website similarity measurements for detecting similar phishing websites using \emph{Normalized Compression Distance} (NCD).
    \item We provide an analysis on the similarities and differences between phishing and legitimate website contents and visual appearances, and how content-based methods would effectively detect phishing attacks better than visual-based methods. 
    \item We propose PhishSim as a tool to effectively detect slightly modified or near-similar phishing websites using prototype-based learning algorithms and the \emph{Normalized Compression Distance}, which is a parameter-free and application independent distance metric to measure similarities between websites' HTML content. This tool works by measuring the pairwise similarity between websites in the dataset, clustering these websites, and performing phishing classifications based on whether a website is grouped in the same cluster with a known phishing website.
    %\item We develop an \emph{unsupervised feature-free} algorithm for classifying phishing websites. This algorithm works by measuring the pairwise similarity between websites in the dataset, clustering these websites, and performing phishing classifications based on whether a website is grouped in the same cluster with a known phishing website.
    %\item We propose the use of a prototype-based learning method with the ability to continuously retain knowledge of known and observed phishing websites, and incrementally learn new forms of phishing websites.
    %\item We propose the use of an \emph{incremental} learning method using phishing website \emph{prototypes} with the ability to continuously retain knowledge of known and observed phishing websites, and incrementally learn new forms of phishing websites.
    \item We introduce a feature-free phishing detection system architecture that can be deployed within intranet servers or in the cloud. 
\end{itemize}

The paper is organized as follows. Section II provides an overview of past studies which are related to our work in phishing detection systems. In section III, we introduce the normalized compression distance (NCD) and some background mathematical concepts which are used in this paper. Section IV describes the overview of our system. Section V provides results on the website similarity analysis to observe the characteristics between a phishing website and its legitimate target website based on the pairwise NCD value. Section VI provides results on the optimal distance threshold selection process. We describe the experimental setup in Section VII, then provide the performance evaluation results in Section VIII and further analysis on these results in Section IX. At the end of the paper, in Section X, we wrap up with conclusions.

\section{Related Works}
\label{sec:related-works}

There is significant research focusing on phishing detection. Some studies focus on the use of blacklists and whitelists in anti-phishing systems. Blacklist-based methods keep a list of domain names or links to known phishing websites and alert users if they try to visit those sites. However, phishing websites are highly dynamic, and the average lifetime of a phishing webpage is only a few hours \cite{aleroud2017phishing}. In many cases, zero-hour phishing attacks easily bypass blacklist-based methods. Meanwhile, whitelist-based approaches allow users to browse only those webpages that are deemed safe, which is often impractical.

Other studies in phishing detection use similarity-based methods to measure similarities between websites by analyzing a website's textual content or screenshot. Text similarity-based methods analyze the semantics of the textual content of emails and webpages to decide whether they classify as a phishing attempt. This method is likely to fail in the future with the increased use of code obfuscation techniques \cite{khonji2013phishing}. A past study by Chen et al. \cite{chen2010detecting} introduced the use of the normalized compression distance (NCD) to measure visual similarities between webpages using the website screenshots. Their work is based on the assumption that phishing websites usually look almost identical to the legitimate website they are targeting. However, this method might not perform well when detecting phishing websites which have a distinct appearance to the target website (Figure~\ref{fig:netflix_screenshots_1} and Figure~\ref{fig:netflix_legitimate}). Cui et al. \cite{cui2017tracking} proposed the proportional distance metric, which computes similarities between websites based on the number of occurrences of a predefined set of HTML tags. A more recent work by Feng et al. \cite{feng2020detection} studied the use of Doc2Vec, which is a deep learning based method to create a numerical representation of phishing websites, and performed phishing website detection based on the similarity of the website's numerical representation.

%Our work focuses on using HTML contents since several phishing websites have identical HTML content while having noticeably different appearance, for instance, Figure~\ref{fig:netflix_screenshots_1}(a) and (b). Many phishing attackers used phishing kits that are available in the dark web, thus several phishing websites tend to have similar HTML content.

% \begin{figure}[H]
%     \centering
%     \subcaptionbox{Legitimate}[0.47\linewidth][c]{\includegraphics[width=\linewidth]{image/L_DBX_0.png}}\quad
%     \subcaptionbox{Phishing}[0.47\linewidth][c]{\includegraphics[width=\linewidth]{image/P_DBX_0.png}}

%     \caption{Dropbox Websites}
%     \label{fig:phish_legit_comparison}
% \end{figure}

Several other studies have attempted to find intrinsic characteristics of phishing websites and use machine learning algorithms to perform classification based on these features. Whittaker et al. \cite{Whittaker2010} introduced a classification system to maintain Google's phishing blacklist automatically based on features extracted from the website's URL, page hosting information, and page content. Furthermore, Zhang et al. \cite{Zhang2007cantina}  developed a framework, called CANTINA, to identify phishing targets using TF-IDF (Term Frequency-Inverse Document Frequency)  analysis and seven other content-based heuristics, including domain age, logo image and domain name inconsistency, as well as suspicious links in the HTML. Purwanto et al. \cite{purwanto2020phishzip} designed a new compression-based algorithm \textsc{PhishZip} to detect phishing websites. Xiang et al. \cite{Xiang2011} improved CANTINA by proposing a more comprehensive framework to detect phishing websites named CANTINA+, which makes use of URL-based, HTML-based, and web-based features. Xiang and Hong \cite{xiang2009hybrid} also proposed a hybrid phishing detection approach by extracting keywords and performing identity discovery with named entity recognition. \color{black}Araujo and Martinez-Rico proposed a phishing detection system which combines the use of website link and language-model (LM)-based features \cite{araujo2010web}.  \color{black}Meanwhile, Zhang et al. \cite{zhang2011textual} introduced a phishing website detection system that analyses the website's textual and visual content, and assesses the similarities. In a recent study, Quinkert et al. \cite{quinkert2019s} attempted to identify scams and phishing by looking for homograph domains that use visually similar Unicode characters to create URLs that look identical.

\color{black}Several past studies focused on using natural language processing (NLP) based techniques to perform phishing website classification. Zhang et al. \cite{zhang2017boosting} proposed the use of text semantic features together with statistical features from the website to improve Chinese phishing website detection. Meanwhile, Opara et al. \cite{opara2020htmlphish} proposed HTMLPhish which uses convolutional neural networks to learn the feature representation of websites based on the semantic dependencies of the website's textual content. The use of semantic analysis has shown to perform well in identifying phishing and legitimate websites, since specific words has shown to occur more frequently in phishing websites \cite{purwanto2020phishzip}, e.g. email, sign, account, password. However, the use of NLP based techniques limits the phishing detection to a specific language.

% add more studies in NCD (and its application in various fields)
% A number of studies have previously discussed the use of data compression algorithm to perform text classification in various areas. Marton et al. in \cite{marton2005compression} evaluated the performance of various compression-based classification methods to classify text based on topic/genre and authorship attribution. Meanwhile, Ziegelmayer and Schrader in \cite{ziegelmayer2012sentiment} discussed the use of Prediction by Partial Matching (PPM) to perform sentiment polarity classification. Compression algorithms have also been used to perform classification of amino acid sequences of DNA as discussed in \cite{chiba2001classification}.

%\section{Assessing Website Similarity from NCD}
\section{Concepts and Definition}
\label{sec:concepts_and_definition}
In this section, we introduce and discuss the use of normalized compression distance (NCD) to perform phishing website detection, and prototype-based learning algorithms for clustering and classifying websites.

\subsection{NCD for Measuring Phishing Website Similarity}
\label{sec:ncd_application_phishing_detection}
NCD is a parameter-free distance measure, which is universal such that it attempts to approximately measure the similarity of dominant features in all pairwise file or object comparisons. The aim of NCD is to capture each effective distance, including the effective versions of Hamming distance, Euclidean distance, and edit distance. \color{black}Further details on NCD is provided in Appendix~\ref{appendix_ncd}\color{black}. The generic characteristics of NCD make it applicable to various kinds of applications \cite{cilibrasi2005clustering}. Based on these characteristics, we have set out to explore whether the use of NCD as a non-feature similarity metric is suitable in the study of phishing detection systems. With the dynamics of phishing, a detection system which relies on a specific and static set of features would potentially fail to detect phishing once the attack behavior changes.

Besides the selection of NCD as the similarity metric, another aspect to consider is the selection of object or data to be compared in the context of phishing detection. Chen et al. \cite{chen2010detecting} use NCD to measure visual similarities between websites and detect phishing websites by calculating the NCD between two website screenshot images. Their study attempts to detect deceptive phishing attacks where the phishing sites are visually very similar to their target legitimate website. Chen et al. argue that while there are some small differences between phishing and legitimate websites, attackers must design the phishing page to be similar to the legitimate page in order to convince users to believe that the website is legitimate. Based on this assumption, they performed phishing website detection by computing the NCD between a suspicious website and the legitimate website. An NCD value below a certain threshold indicated that the website was indeed imitating the legitimate website, therefore categorizing it as a phishing website. While this assumption is true for some phishing websites, we found that in many other cases, phishing websites are not necessarily identical to their target website.

On the other hand, Cui et al. \cite{cui2017tracking} demonstrated that 90\% of the phishing data collected in a 10-month period in 2016 were variations or replications of other previous  attacks in the database, which indicates that there are repetitions and similarities among the phishing websites themselves. This is also understandable as there has been an increase in the use of phishing kits which increase the possibility of similar HTML contents from new phishing websites. Thus, we take a new approach where we  perform phishing detection by detecting similarities in the website HTML content, as phishing websites are often developed from a certain template or kit. Therefore, in this study, we perform pairwise NCD calculations between website HTML datasets to measure website similarities and detect phishing websites. Further details regarding the system design will be discussed in the next section. % ref to the relevant section

\subsection{Prototype-based Learning}
\label{sec:cluster_and_classification}
Using NCD metrics to measure the similarities between two websites, we perform \textit{clustering} to  divide phishing websites with similar HTML contents into a number of groups and \textit{classification} to assign a website to the previously generated group with the closest similarity. The aim of our clustering is to group similar websites, indicated by the relatively small number of pairwise NCD values.

\begin{figure}
    \centering
    \includegraphics[width=0.5\linewidth]{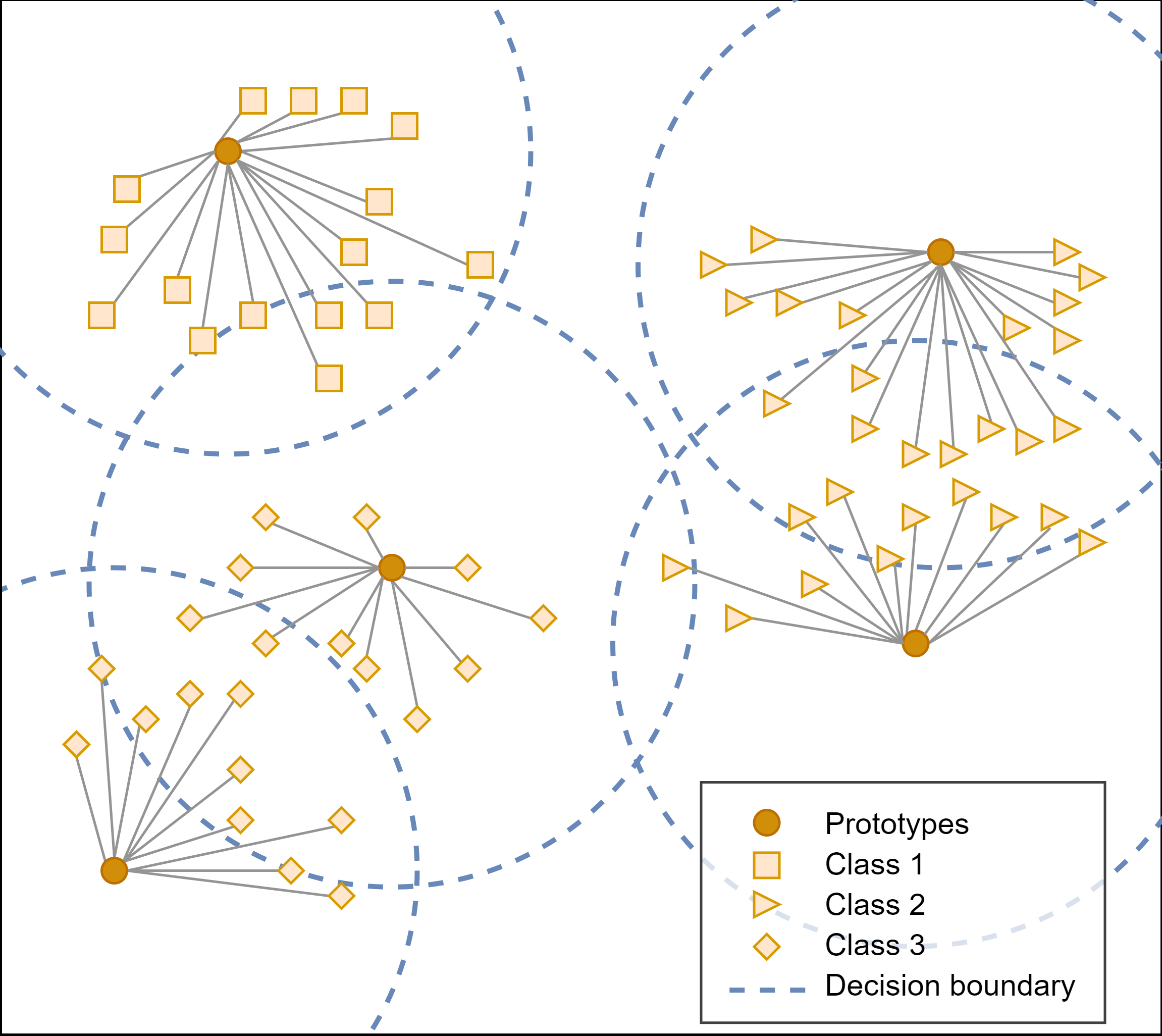}
    
    \caption{Classification using Prototypes}
    \label{fig:prototypes}
\end{figure}

When deciding which approach to take when performing clustering and classification, we take into consideration the main goals or characteristics which are important to have for the detection to be accurate and sustainable. Firstly, we aim for the system to be feature-free, meaning that it can learn directly from the data without the need to perform manual feature extraction. Furthermore, one of the advantages of using a feature-free approach is that the system can adapt to changes in phishing behavior or data representation. Secondly, we attempt to build a detection system that has the ability to continuously learn and gain knowledge from the previous learning process, and incorporate this with new knowledge obtained while processing new data, resulting in a continuously learning system. With the ability to incrementally learn, the goal is that the system will be able to improve its detection over time as it meets new phishing examples.

For the reasons mentioned above, a prototype-based clustering approach has been selected, where clusters are represented by actual samples in the dataset instead of centroids. By using actual website samples as cluster representations, we are able to implement an incremental feature-free learning method as the prototype-based method does not necessarily require the data to be transformed into a specific feature representation. The prototype-based method fits our purpose of designing a fast, efficient, and feature-free algorithm. Using a greedy clustering algorithm \cite{cui2017tracking}, the overall time-complexity for $n$ data samples would be $O(n^2)$, whereas it would only be $O(nk)$ using the prototype-based approach, where $k$ is the number of clusters \cite{gonzalez1985clustering}.

By definition, a \textit{prototype} is a data point, which represents all the data points in a cluster. Each data point in a given cluster can be assigned to a specific prototype, where the distance from this prototype to the data instance is less than a certain  threshold  distance $d$. Each prototype in our detection system would be one or several phishing websites which represent a cluster of similar phishing websites, potentially generated using the same template or phishing kit. However, we are not focusing on which cluster or class of phishing websites each dataset belongs to. Instead, we aim to detect whether a website would be categorized as one of the phishing classes or whether it is not similar to any class at all. Therefore, our detection approach attempts to check whether a website is similar to any phishing cluster/class, by measuring the pairwise NCD values between the website and each prototype. Further details on the algorithms will be given in the following sections.

There is no specific requirement as to what distance metrics to choose (e.g., Hamming distance, Euclidean distance, and edit distance).  However, in our case, we have selected NCD as the distance measure for reasons that were mentioned in the earlier sections. The concept of prototypes is illustrated in Figure~\ref{fig:prototypes}. In this figure, roughly 60 data instances are grouped in three classes, which are represented using five prototypes. In some cases, a cluster can also be represented by more than one prototype, e.g., the clusters in  the left-bottom and right parts of Figure~\ref{fig:prototypes}. Using the prototypes, we are also able to perform classification on a new data point $x$ by checking if the data is similar to one of the prototypes, i.e., $NCD(x,z) < d_{threshold}$ for each $z \in prototypes$ (the set of prototypes). As seen in Figure~\ref{fig:prototypes}, it is possible that the new data sample may be close to two or more prototypes, or located in the regions where the prototype decision boundaries overlap with one another. For simplicity, we have assigned such data samples to the prototype which is the closest (i.e., with the smallest NCD). Note that the clusters are circular in nature, with a prototype at the center. This is in contrast with Lloyd's algorithm for the $k$-means clustering problem, which produces Voronoi cells, with the cluster center as the centroid. %Further details on the clustering and classification algorithms are provided in Appendix~\ref{appendix_prototype}

\begin{figure}
    \centering
    \subcaptionbox{\scriptsize First prototype}[0.28\linewidth][c]{
    \includegraphics[width=\linewidth]{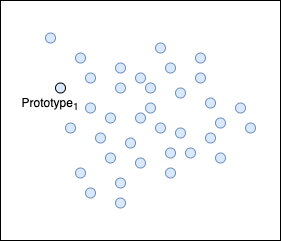}}\quad
    \subcaptionbox{\scriptsize Second prototype (furthest point from $prototype_1$)}[0.28\linewidth][c]{
    \includegraphics[width=\linewidth]{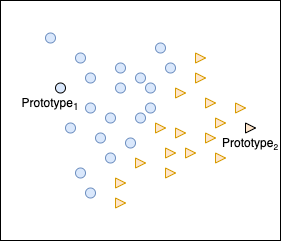}}\quad
    \subcaptionbox{\scriptsize Third prototype (furthest point from $prototype_2$)}[0.28\linewidth][c]{
    \includegraphics[width=\linewidth]{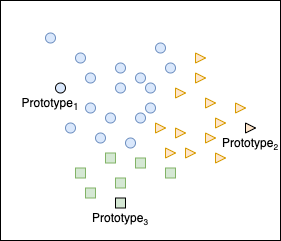}}
    \hfill
    \caption{Furthest Point First Algorithm}
    \label{fig:fpf}
\end{figure}

% The prototype extraction algorithm we use in our proposed detection method is provided in Algorithm~\ref{alg:prototype_extraction} (Appendix~\ref{appendix_prototype}). 

\subsubsection{Prototype Extraction}
\label{sec:prototype_extraction}
\color{black}Extracting representative prototypes from the data is a non-trivial task. The algorithms used to perform prototype extraction is inspired by \cite{gonzalez1985clustering}, with changes made to use NCD as the distance function.\color{black} The algorithm adopted an $O(nk)$-time algorithm originally proposed by Gonz\`alez \cite{gonzalez1985clustering}, in which extracting $k$ prototypes from $n$ data points takes an objective function value at most twice the size of the optimal solution. Gonz\`alez's Furthest Point First (FPF) algorithm is illustrated in Figure~\ref{fig:fpf}. The main idea behind this algorithm is to select a data instance as a prototype, assign all data to the closest prototype, then choose the next instance furthest from the current prototype as the next prototype. This process is repeated until all data is assigned to a prototype with NCD values less than $d_{threshold}$. After performing prototype extraction by running this algorithm, we will have a set of prototypes representative of the phishing data.
\color{black}

% By running Algorithm~\ref{alg:prototype_extraction}, we can extract the prototypes which are representative of the phishing websites. 

\subsubsection{Incremental Learning}
\label{sec:incremental_learning_algo}
For continuous phishing website detection, the use of an incremental learning algorithm is proposed, which is adapted from the incremental learning algorithm introduced by Rieck et al. \cite{rieck2011automatic}. The main idea is to perform NCD-based classification in each iteration, and update the set of prototypes with new prototypes extracted from misclassified phishing samples belonging to new data in every iteration.

\subsubsection{Classification}
\label{sec:proto_classification}
With the list of prototypes, we are able to classify whether a website is similar to one of the known phishing websites. This is done by computing the NCD between the website $x$ and each phishing website prototype $z$, and checking whether a prototype exists such that the NCD value is less than a certain maximum distance value or $NCD(x,z) < d_{threshold}$. \color{black}The algorithm to perform classification for this work is adapted from the method proposed by Rieck at al. \cite{rieck2011automatic}, as shown in Algorithm~\ref{alg:ncd_classification}.\color{black}

\begin{algorithm}[t]
\caption{Prototype Extraction}
\label{alg:prototype_extraction}
\begin{algorithmic}[1]
\State $prototypes \leftarrow \emptyset$

\ForAll {$x \in data$}
    \State $distance[x] \leftarrow \infty$
    \State $cluster[x] \leftarrow \emptyset$
\EndFor

\While {$max(distance) > d_{threshold}$}
    \State $z \leftarrow \argmax_{x \in data} distance[x]$
    \For {$x \in data$}
        \If {$distance[x] > NCD(x,z)$}
            \State $distance[x] \leftarrow NCD(x,z)$
            \State $cluster[x] \leftarrow z$
        \EndIf
    \EndFor
    \State $prototypes \leftarrow prototypes \cup \{z\}$
    \State $data \leftarrow data \setminus \{z\}$
\EndWhile

\end{algorithmic}
\end{algorithm}

% This algorithm is provided in Algorithm~\ref{alg:ncd_classification} (Appendix~\ref{appendix_prototype}).

\begin{algorithm}[t]
\caption{NCD-based Classification}
\label{alg:ncd_classification}
\begin{algorithmic}[1]

\For {$x \in data_{test} $}
    \State $z \leftarrow \argmin_{p \in prototypes} NCD(x,p)$
    \If {$NCD(x,z) < d_{threshold}$}
        \State classify $x$ as phishing
    \Else
        \State classify $x$ as non-phishing
    \EndIf
\EndFor

\end{algorithmic}
\end{algorithm}

\begin{algorithm}[t]
\caption{Incremental Learning}
\label{alg:incremental_learning}
\begin{algorithmic}[1]
    \For {$data \leftarrow source$}
        \For {$x \in data$}
            \State $\text{classify } x \text{ using } prototypes$ \Comment{Algorithm~\ref{alg:ncd_classification}}
        \EndFor
        \State $rejected \leftarrow \text{samples in } data \text{ rejected as phishing}$
        \State $prototypes_{new} \leftarrow \text{prototypes in } rejected$
        \Comment{Algorithm~\ref{alg:prototype_extraction}}
        \State $prototypes \leftarrow prototypes \cup prototypes_{new}$
    \EndFor
\end{algorithmic}
\end{algorithm}

% The incremental learning algorithm is provided in Algorithm~\ref{alg:incremental_learning} (Appendix~\ref{appendix_prototype}). 

% (Algorithm~\ref{alg:prototype_extraction})

\subsection{Optimal Distance Threshold}
\label{sec:optimal_threshold}
When performing prototype extraction, we need to select the maximum distance threshold that defines the size of each prototype's covering, which in the end will affect the size of each cluster and the classification's performance. A common and intuitive characteristic of a good clustering threshold is one that produces compact or dense clusters, which are far away from each other \cite{cui2017tracking}. Based on these characteristics, we define a \emph{quality of clustering (QC)} metric, which is obtained by calculating the ratio between the average cluster compactness and the minimum distance between the prototypes, as shown in Equation~\ref{eq:qc}. The \emph{compactness} of each cluster is computed by taking the average of pairwise NCD values between each data instance in a cluster with the corresponding prototype (Equation~\ref{eq:cluster_compactness}). Note that the more compact the cluster, the smaller this value will be. \emph{Minimum inter-cluster distance (MICD)} is the minimum NCD value between every combination of prototype pairs (Equation~\ref{eq:proto_distance}). The further apart the clusters are, the larger this value will be. Therefore, a smaller QC value represents a better quality of clustering.

\begin{equation}
\label{eq:cluster_compactness}
    Compactness(C_i)=\frac{1}{|C_i|}\sum_{x \in C_i} NCD(x,prototype_i)
\end{equation}

\begin{equation}
\label{eq:proto_distance}
        MICD = \min_{x,y \in prototypes} NCD(x,y)
\end{equation}

\begin{equation}
\label{eq:qc}
    QC = \frac{\frac{1}{|C|} \sum_{i=1}^{|C|} Compactness(C_i)}{MICD}
\end{equation}

During the threshold selection process, our aim is to select the distance threshold $d_{threshold}$ which leads to the best clustering quality. As smaller QC implies better cluster quality, the threshold selection process is done by  performing optimization to obtain the minimum QC parameter.

\section{PhishSim System Overview}
\label{sec:system-overview}
\begin{figure}
    \centering
    \includegraphics[width=0.6\linewidth]{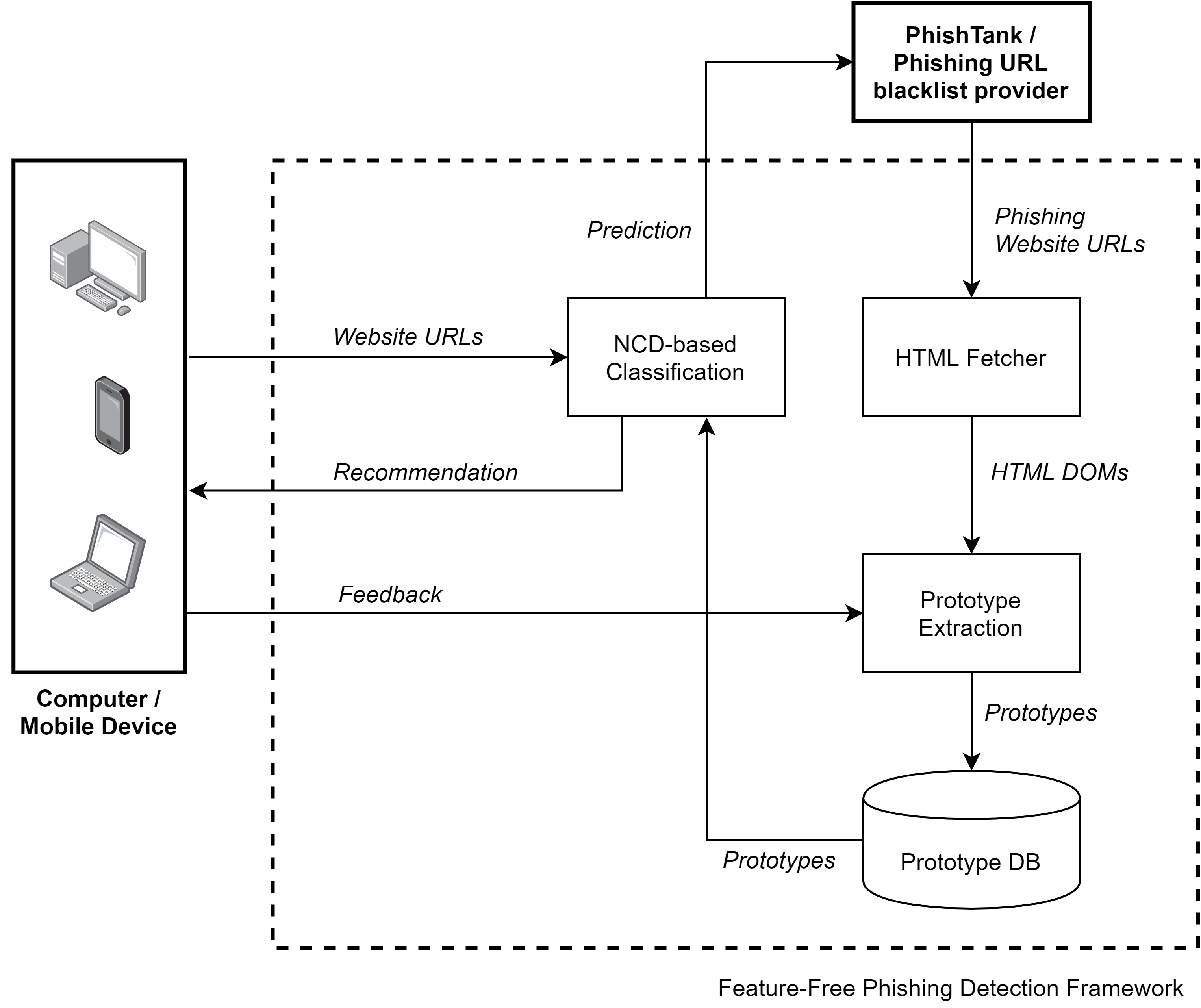}
    \caption{System Diagram of the Feature-free Phishing Detection Framework \emph{PhishSim}}
    \label{fig:system_diagram}
\end{figure}

We propose a server-based phishing detection system \emph{PhishSim} that can be deployed by enterprises in their corporate intranets, Internet Service Providers (ISP), and cloud providers like Amazon, Microsoft, and Google, to defend against phishing attacks. Figure~\ref{fig:system_diagram} shows a high-level overview of how the system works and how it can be implemented. The system receives website URLs requested by users as input. Then, it provides recommendations about whether the webpages are safe or malicious. The recommendation output is generated from the NCD-based classifier using prototypes stored in its \emph{Prototype DB} database. This system also has an ability to update its prototype database by receiving a new phishing website list and generating new prototypes. The new phishing websites can be obtained from user reports and feedback, and from phishing blacklist providers, such as PhishTank \cite{phishTank}.

In the following subsections, we describe the two main components in our proposed feature-free phishing detection system: \emph{phishing website classification} and \emph{phishing prototype database update}.

\subsection{Phishing Website Classification}
To perform phishing website detection, our proposed system receives the website URL when a user is about to open the website's HTML Document Object Model (DOM), hence simulating how the page is rendered by a web browser. To obtain the website's HTML DOM, we used Chromium \cite{chromium}, an open-source software project, which is the basis of several web browsers, including Google Chrome and Microsoft Edge.

% (Algorithm~\ref{alg:ncd_classification})

After obtaining the website's HTML DOM, the system performs classification by measuring NCD-based similarities between the website and known phishing prototypes in the database. The classification process returns a prediction on whether the website is phishing or legitimate. If the website is suspected to be a phishing website, this result will be added to the prototype database or blacklist provider, and users will be redirected to a warning page (with a recommendation to not open the page) when accessing the webpage. Our system can be used in conjunction with Google Safe Browsing \cite{Whittaker2010} that is based on the list of URLs for websites that contain phishing content.

To enhance the robustness and prevent attackers from evading detection, we removed the text and HTML comments in the content prior to performing compression, leaving only the HTML tags which are rendered and visually shown on the browser. Thus, the addition of invisible elements in the HTML will not affect the performance. 

% (Algorithm~\ref{alg:prototype_extraction}) 

\subsection{Phishing Prototype Database Update}
To keep the system prediction accurate, our system has a mechanism to frequently update the phishing prototype database. The system is able to update its prototype database by periodically receiving data from phishing blacklist providers (e.g., daily or weekly)  and users' reports. After obtaining the websites' HTML DOMs, the system performs prototype extraction to extract representative prototypes from these new data. Afterwards, the extracted prototypes are stored in the prototype database used for NCD-based classification.

\section{Similarity Analysis}
\label{sec:similarity_analysis}
To validate our method, we conducted two experiments to observe the characteristics and similarities between a phishing website and its legitimate target website by performing pairwise NCD computations. In the first experiment, we performed pairwise NCD calculations and clustering on data targeting a specific brand, and we focused on observing the similarities between phishing and legitimate websites. The aim of this experiment was to observe the relationship between each website, specifically the relationship between websites (phishing and legitimate) of different brands. In the second experiment, we applied the same approach as in the second experiment to phishing and legitimate website data of multiple brands.

% In the first experiment, we observed how minor image data alterations (e.g., pixel-shifting and colour changes) would affect the NCD values by performing analyses on artificial data. The purpose of this experiment was to observe the robustness of the use of NCD to detect visually similar websites. 

The phishing data used in this experiment is from websites reported by Internet users to \emph{PhishTank} \cite{phishTank} and collected between March and April 2020. To identify the target brand of each phishing website, we utilized the target brand information, which was provided by PhishTank. However, we noticed that many phishing websites reported to PhishTank did not provide target brand information (i.e., identified as 'Others'). Thus, we also made use of Google's Cloud Vision API to estimate the target brand by detecting the logos that appeared in website screenshots. Lastly, we performed manual checking after grouping the data based on its brand to ensure that the target brand was correct.

\subsection{Brand Specific Similarity Analysis}
\label{sec:brand_specific_analysis}
For comparison with Chen et al. \cite{chen2010detecting}, we also performed website similarity analyses using the website screenshots. In this experiment, we focus on three brands, which are Microsoft, PayPal, and Netflix, as we found that these brands dominate our phishing website data.  Furthermore, VadeSecure reported that these brands were among the top five phishing targets in 2019 and the top three phishing targets in Q1 and Q3 2019 \cite{vadesecure2019pishersFavoritesQ3, vadesecure2019pishersFavoritesQ1, vadesecure2019pishersFavoritesQ2, vadesecure2019pishersFavoritesQ4}. For each brand, we included the legitimate website data and ten corresponding phishing websites selected at random.

We first performed pairwise NCD computation using HTML DOM files to observe whether similarities exist among the phishing websites, and whether they are also similar to the legitimate target website. To gain a better understanding of the relationships between the websites, we performed data clustering on the pairwise NCD data and visualized the data as dendrograms. In the dendrograms provided in the following subsections, we followed a strict naming convention with three label codes (separated by an underscore). The first code (prefix), indicates whether it is a legitimate or phishing website ('L' for legitimate, 'P' for phishing). The second code indicates the brand ('MST' for Microsoft, 'NTF' for Netflix, and 'PYL' for PayPal). The third code (suffix) indicates the website number. In the following subsections, we will focus our discussion on the findings for each brand.

% \begin{figure}
%     \centering
%     \includegraphics[width=0.485\textwidth]{image/mini_experiment_results/netflix_html_dist_matrix.png}
%     \caption{Netflix Distance Matrix (HTML DOM)}
%     \label{fig:dist_matrix_netflix_html}
% \end{figure}

\begin{figure}
    \centering
    \includegraphics[width=0.8\linewidth]{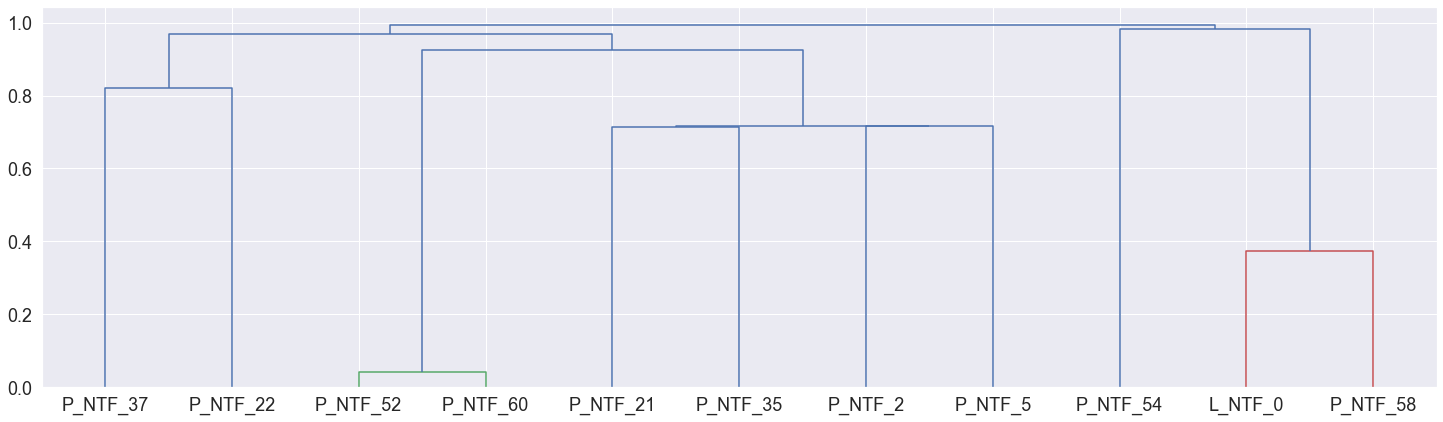}
    \caption{Netflix dendrogram (HTML DOM)}
    \label{fig:dendogram_netflix_html}
\end{figure}

\begin{figure}
    \centering
    \includegraphics[width=0.8\linewidth]{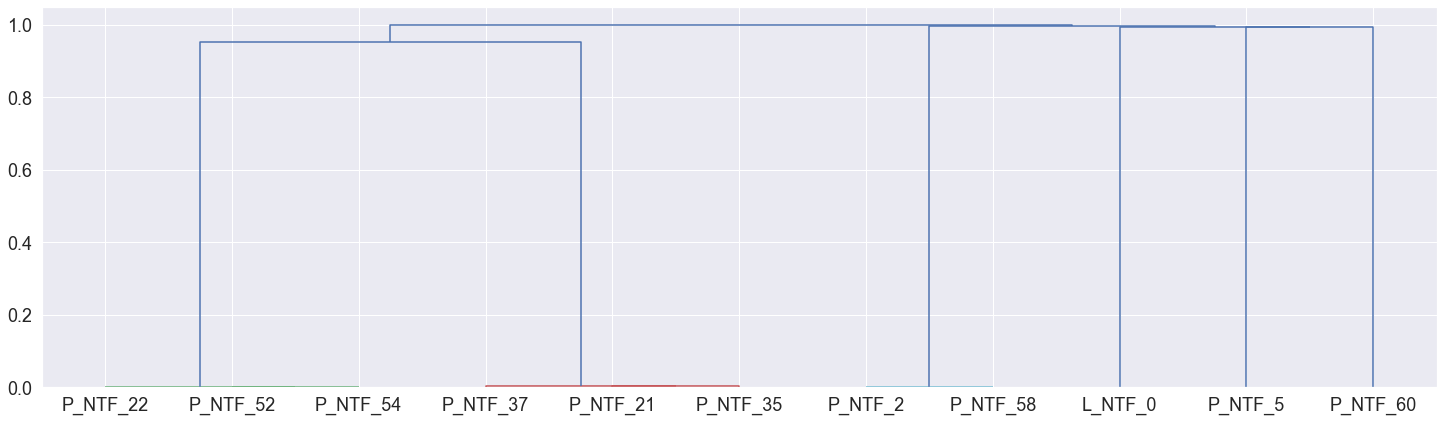}
    \caption{Netflix Dendrogram (Website Screenshot Image)}
    \label{fig:dendogram_netflix_screenshot}
\end{figure}

In this section, we will only include similarity analyses on Netflix phishing and legitimate websites, as we found that the design of the Netflix phishing websites has more variations. These variations are potentially due to frequent changes of the design of the target legitimate website itself.

% Meanwhile, the similarity analyses on Microsoft and PayPal phishing and legitimate websites are provided in Appendix~\ref{appendix_brand_specific}.

% \begin{figure}
%     \centering
%     \includegraphics[width=0.45\textwidth]{image/mini_experiment_results/L_NTF_0.png}
%     \caption{Netflix Legitimate Website}
%     \label{fig:netflix_legitimate}
% \end{figure}

\begin{figure}[t]
    \centering
    \subcaptionbox{P\_NTF\_52}[0.45\linewidth][c]{
        \includegraphics[width=\linewidth]{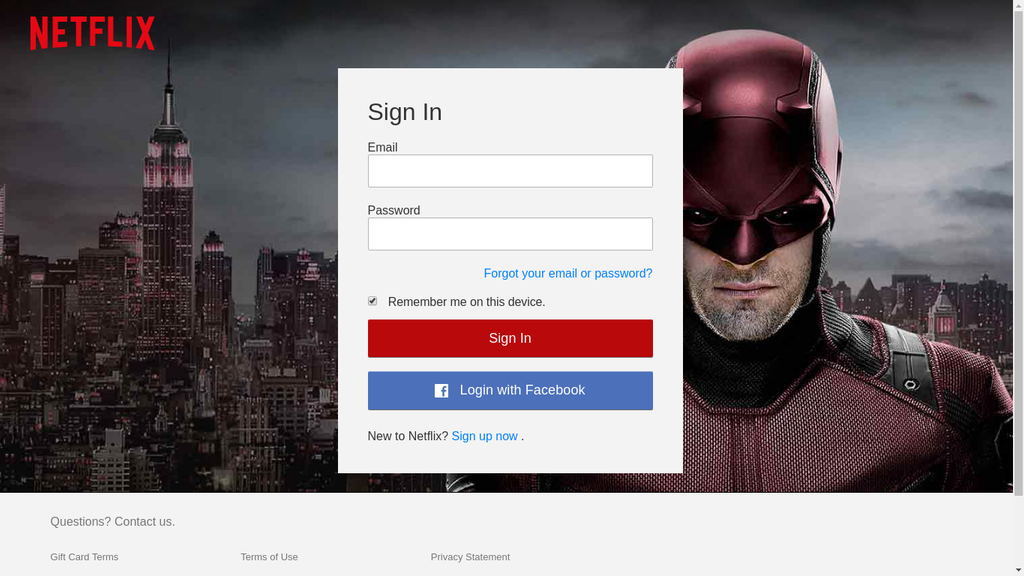}}\quad
    \subcaptionbox{P\_NTF\_60}[0.45\linewidth][c]{
        \includegraphics[width=\linewidth]{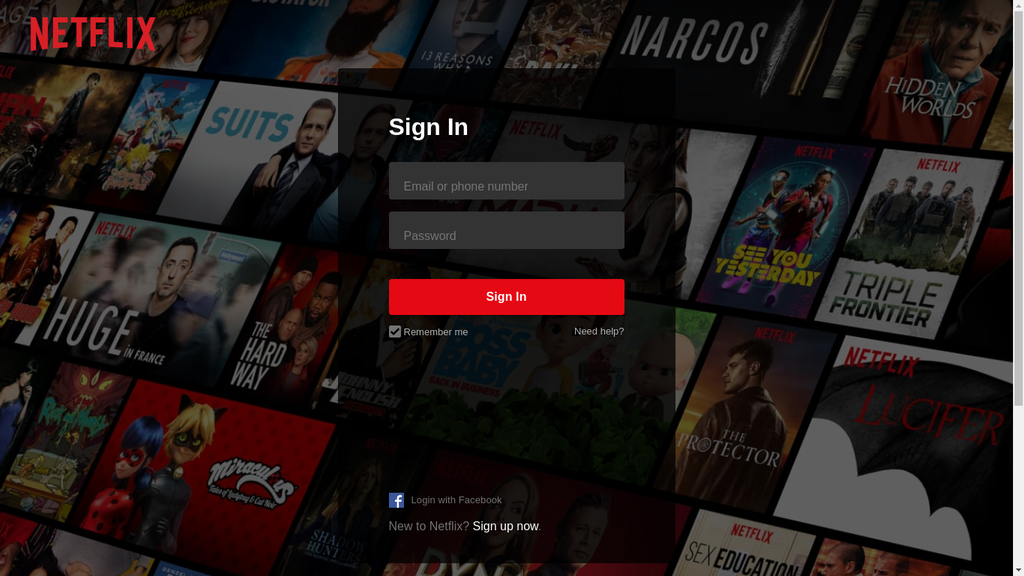}}

    \hfill
    \caption{Phishing Websites with Similar HTML Contents (Cluster 1)}
    \label{fig:netflix_screenshots_1}
\end{figure}

To observe the relationship between these websites based on their content, we compute the pairwise NCD values and perform data clustering on the HTML DOM files. The pairwise NCD values between the HTML DOM files are visualized as a cluster dendrogram as shown in Figure~\ref{fig:dendogram_netflix_html}. Cutting the dendrogram on the highest level gives two clusters. The legitimate website is clustered together with two of the phishing websites, while the remaining phishing websites are clustered together. As shown in the dendrogram, the phishing websites that are most similar to each other are P\_NTF\_52 and P\_NTF\_60, with a NCD of 0.04. Interestingly, these two websites have a very distinctive design, as shown in Figure~\ref{fig:netflix_screenshots_1}. By looking into the HTML DOM files, however, these websites' HTML DOMs are almost identical. The only difference is in a numeric variable set when loading the CSS stylesheet. It is very likely that these websites are built using the same phishing kit, indicated by the similarities in the HTML DOM structure. We also found that this is the case in the other cluster of websites, which are (P\_NTF\_21, P\_NTF\_35, P\_NTF\_2, P\_NTF\_5), and (P\_NTF\_37, P\_NTF\_22). Similar to the first case of P\_NTF\_52 and P\_NTF\_60, the websites which are clustered together have an identical HTML DOM structure, but with different website designs. We also found that in some cases, the websites are similar in style but have different background images. Meanwhile, we also see that the legitimate website has a close relationship with one of the phishing websites, P\_NTF\_58. While this might introduce false positives in the detection system for detecting phishing using NCD-based similarity measurements, we argue that this would be a rare case and it can be avoided by whitelisting legitimate websites. The aim of the whitelist method is to filter legitimate websites by comparing whether the content and domain of the website is identical to one of the whitelisted websites. Using this method, the phishing website will fail to satisfy the properties needed to escape filtering, and would be processed further with the NCD-based similarity assessment.

%However, the implementations of using a whitelist are beyond the scope of this paper.

% \begin{figure}
%     \centering
%     \includegraphics[width=0.485\textwidth]{image/mini_experiment_results/netflix_screenshot_dist_matrix.png}
%     \caption{Netflix Distance Matrix (Website Screenshot Image)}
%     \label{fig:dist_matrix_netflix_screenshot}
% \end{figure}

% \begin{figure}
%     \centering
%     \subcaptionbox{P\_NTF\_21 \& P\_NTF\_35}[0.45\linewidth][c]{
%         \includegraphics[width=\linewidth]{image/mini_experiment_results/P_NTF_21.png}}
        
%     \bigskip
    
%     \subcaptionbox{P\_NTF\_2}[0.45\linewidth][c]{
%         \includegraphics[width=\linewidth]{image/mini_experiment_results/P_NTF_2.png}}\quad
%     \subcaptionbox{P\_NTF\_5}[0.45\linewidth][c]{
%         \includegraphics[width=\linewidth]{image/mini_experiment_results/P_NTF_5.png}}
        
%     \hfill
%     \caption{Phishing Websites with Similar HTML Contents (Cluster 2)}
%     \label{fig:netflix_screenshots_2}
% \end{figure}

\begin{figure}
    \centering
    \includegraphics[width=0.5\linewidth]{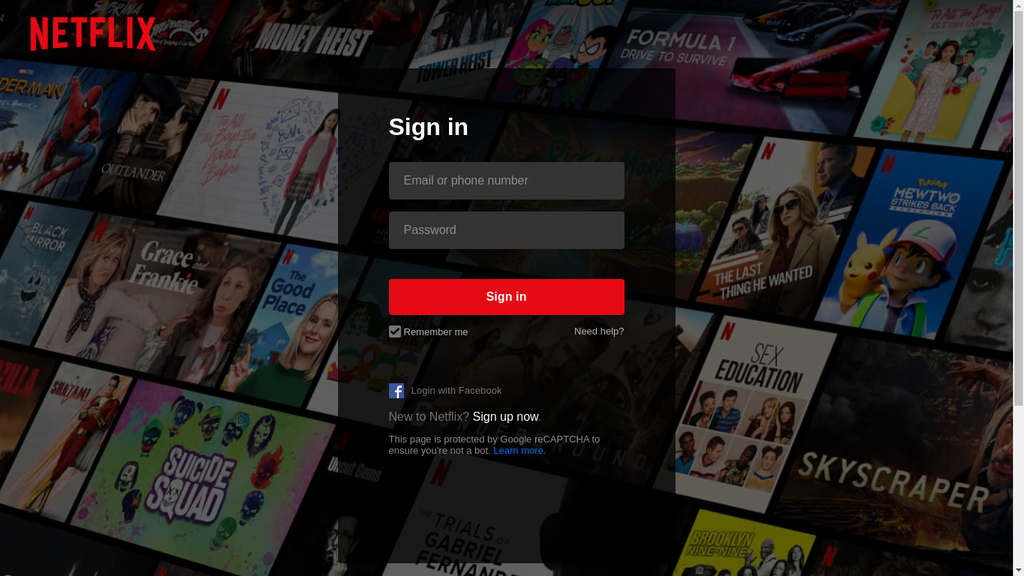}
    \caption{Netflix Legitimate Website}
    \label{fig:netflix_legitimate}
\end{figure}

\begin{figure}
    \centering
    \includegraphics[width=0.8\linewidth]{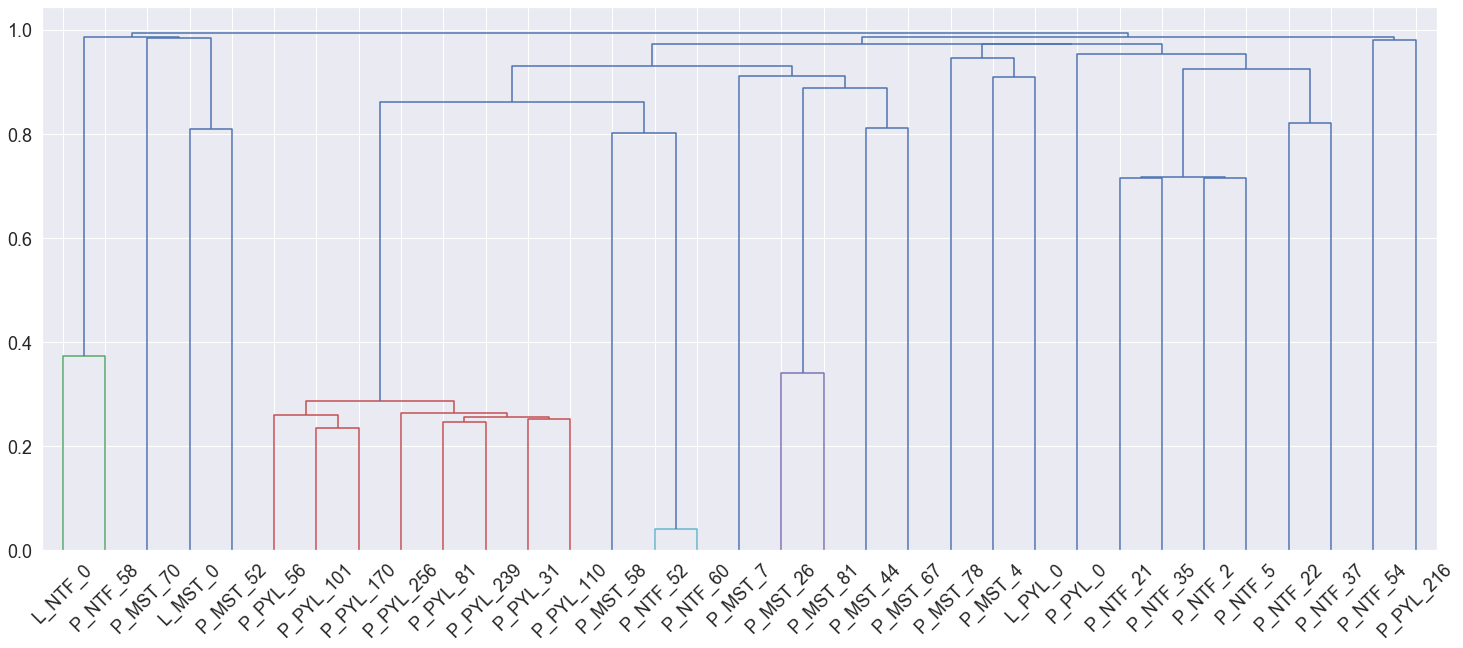}
    \caption{Multi-Brand Website Data (HTML DOM)}
    \label{fig:multibrand_html}
\end{figure}

To observe the relationship between the websites based on their screenshots, we also analyzed the pairwise NCD values between the screenshots. Using the same approach as the Netflix HTML DOM files, we performed clustering with the screenshot image data. The cluster dendrogram is shown in Figure~\ref{fig:dendogram_netflix_screenshot}. Based on these results, it is demonstrated that websites with different designs have pairwise NCD values between 0.95 and 1. Meanwhile, the websites that are detected as having high similarity are clustered together, as shown in the dendrogram. These websites have identical designs, such as (P\_NTF\_22, P\_NTF\_52, P\_NTF\_54), (P\_NTF\_37, P\_NTF\_21, P\_NTF\_35), and (P\_NTF\_2, P\_NTF\_58). Interestingly, the legitimate website (Figure~\ref{fig:netflix_legitimate}) was not detected as similar to any of the phishing websites, despite having similar element styles. The only difference is the background image, which may be updated frequently with images from the latest movie or TV series. Based on this experiment, we were unable to detect website similarities by computing the pairwise NCD values on the website screenshots.

\subsection{Multi-Brand Similarity Analysis}
\label{sec:multibrand_analysis}
\begin{figure}
    \centering
    \includegraphics[width=0.8\linewidth]{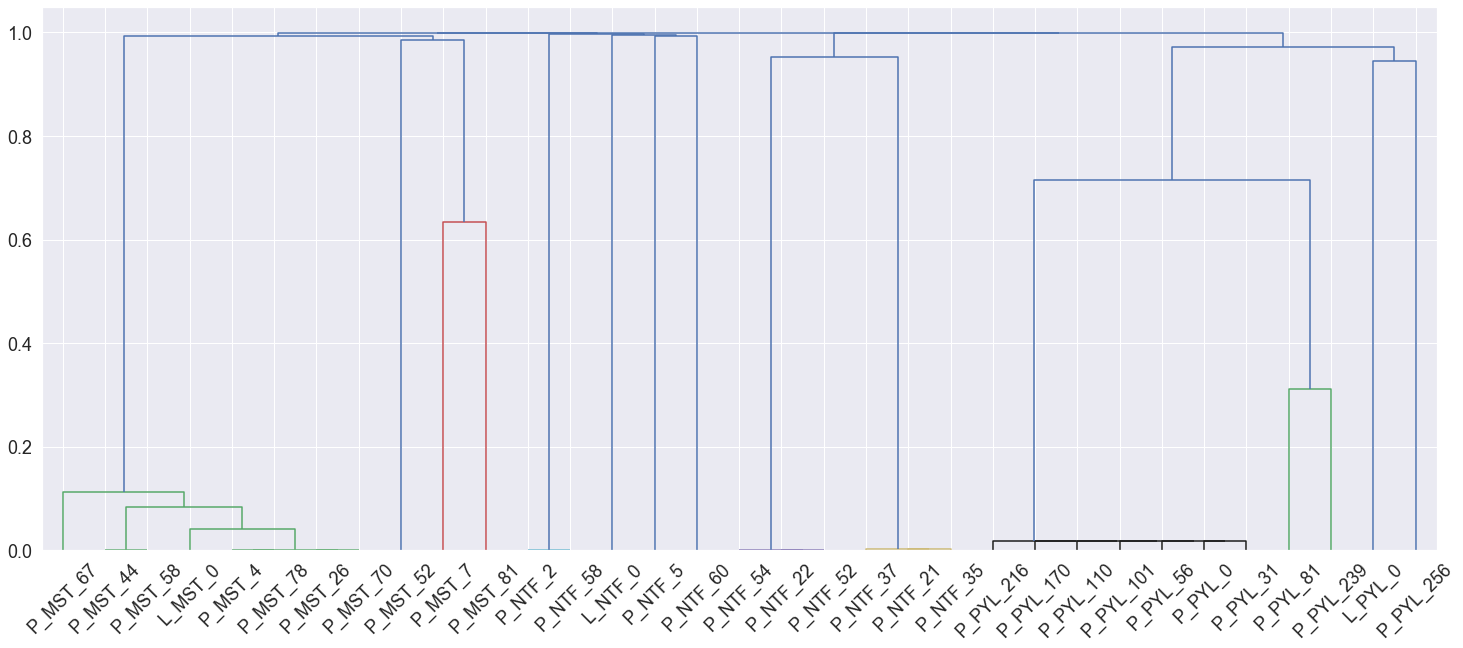}
    \caption{Multi-Brand Website Data (Website Screenshot Image)}
    \label{fig:multibrand_screenshot}
\end{figure}

To observe the relationship between various brands' phishing and legitimate websites, we attempted to compute the pairwise NCD values between the websites. We did this in terms of the HTML DOM and website screenshot images. The brands that we included in this analysis were Microsoft, Netflix, and PayPal, which are the three brands that consistently ranked among the top five phishing targets in every quarter in 2019 \cite{vadesecure2019pishersFavoritesQ3, vadesecure2019pishersFavoritesQ1, vadesecure2019pishersFavoritesQ2, vadesecure2019pishersFavoritesQ4}. Having these pairwise NCD values, we were able to cluster the websites and observe the relationship by visualizing the clusters as a dendrogram. The dendrogram of websites from the pairwise NCD values between the HTML DOM files is shown in Figure~\ref{fig:multibrand_html}, while the dendrogram constructed using the screenshot image files is shown in Figure~\ref{fig:multibrand_screenshot}.

By comparing the dendrograms in Figure~\ref{fig:multibrand_html} and Figure~\ref{fig:multibrand_screenshot}, we are able to capture similarities and differences between the website clusters constructed using HTML DOM files and website screenshot image files. In terms of similarities, phishing websites targeting similar brands are shown to be clustered together and have a close relationship in both dendrograms. However, the first dendrogram (constructed using the HTML DOM files) demonstrates a greater separation between the cluster of phishing websites targeting a certain brand and the cluster where the legitimate target website belongs. Meanwhile, in the second dendrogram (constructed using the website screenshot images), clusters of the same brands tend to be alongside each other. Furthermore, there seems to be very limited separation between the legitimate and phishing websites.

\begin{figure}[t]
    \centering
    \includegraphics[width=0.9\linewidth]{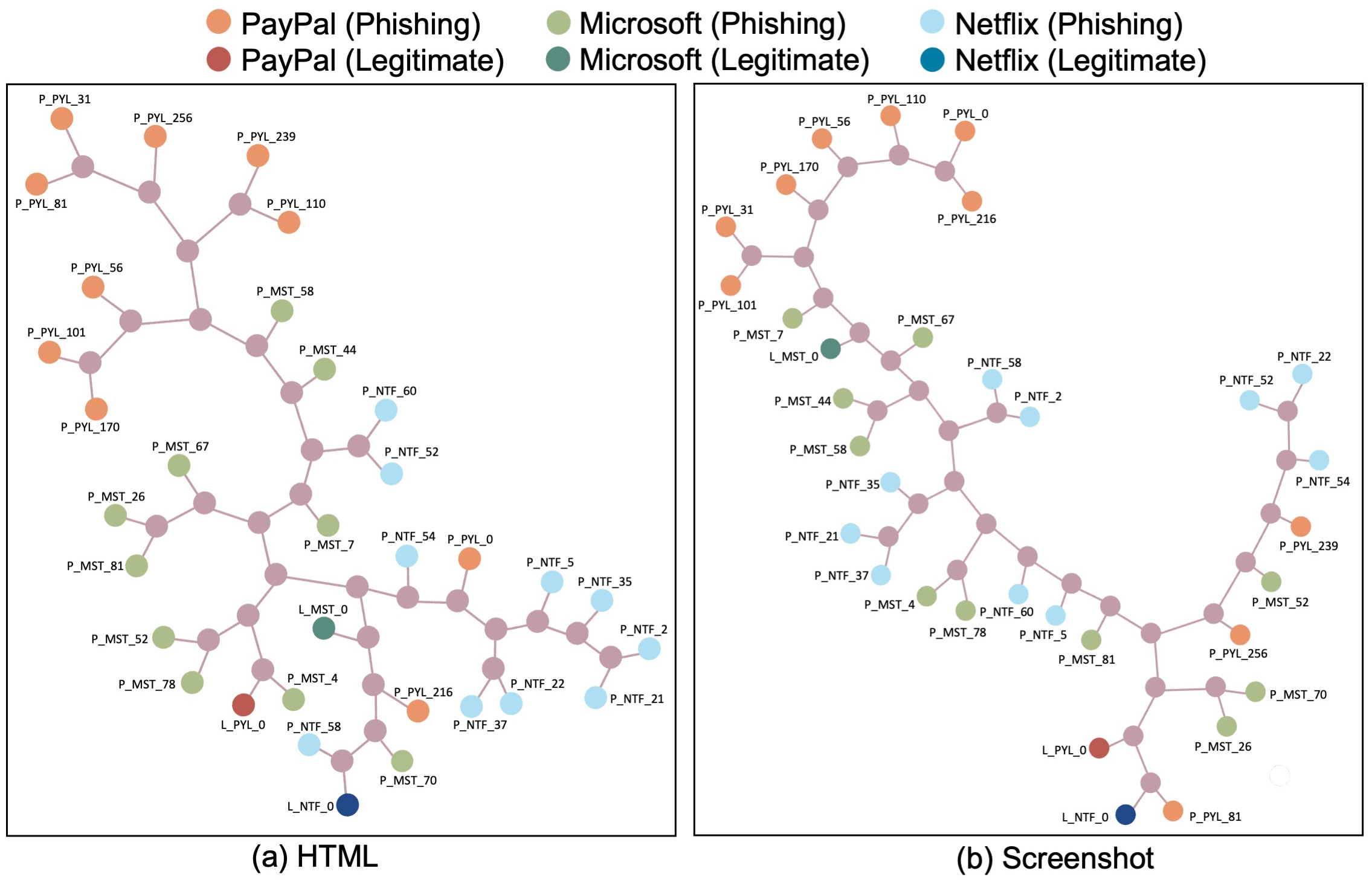}
    \caption{Website Clusters}
    \label{fig:quartet_tree}
\end{figure}

\begin{figure}
    \centering
    \includegraphics[width=0.7\linewidth]{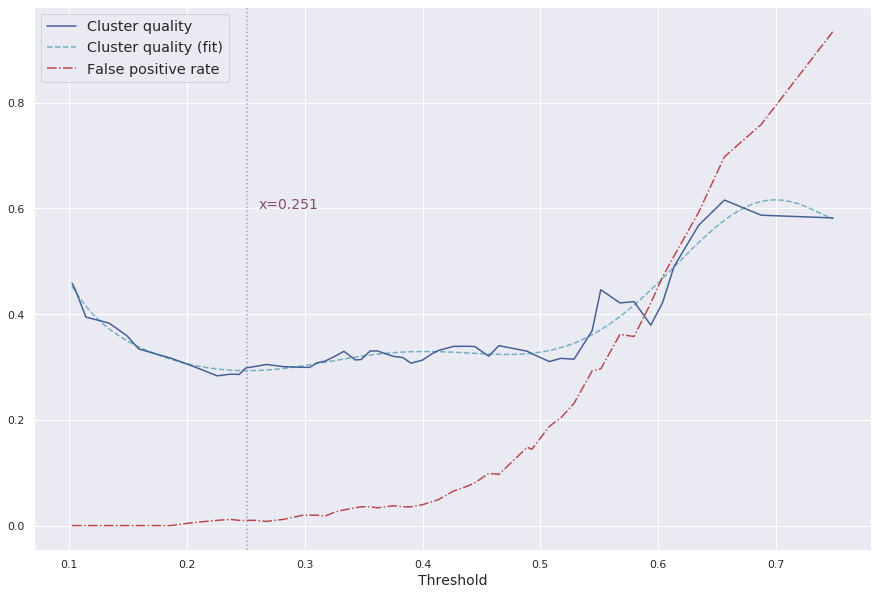}
    \caption{Threshold Selection}
    \label{fig:threshold_selection}
\end{figure}

\color{black}The relationship between each website can also be observed by plotting the clusters as a quartet tree (Figure~\ref{fig:quartet_tree}), which visualizes the closeness of the websites based on their distances to each other. To generate the quartet trees, we used the CompLearn toolkit\footnote{https://complearn.org/index.html}, which uses the Minimum Cost Quartet Tree Reconstruction method to generate the optimal quartet trees that represent the clusters. As shown in Figure~\ref{fig:quartet_tree}, the majority of PayPal phishing websites occupy a single branch on the tree when clustering using HTML and screenshot data. Using the HTML data, it is shown that the Microsoft phishing websites are clustered more closely compared to clustering using screenshot data. The Netflix phishing websites are grouped together in the lower right part of the tree when clustering using HTML data. Meanwhile, using screenshot data, the Netflix phishing websites with identical designs are clustered together (i.e., (P\_NTF\_22, P\_NTF\_52, P\_NTF\_54), (P\_NTF\_37, P\_NTF\_21, P\_NTF\_35), and (P\_NTF\_2, P\_NTF\_58)) as also shown in the dendrogram in Figure~\ref{fig:dendogram_netflix_screenshot}. Based on these quartet trees, the phishing websites of the same brands are relatively closer to each other when clustered using its HTML data compared to the screenshot data, which indicates high similarity between phishing websites' HTML content despite its visual appearance.
\color{black}

\section{Optimal Distance Threshold Selection}
\label{threshold_selection}
We attempted to vary the maximum distance value to obtain the minimum Quality of Clustering (QC) value as defined in Section~\ref{sec:optimal_threshold}. Note that a smaller QC value indicates superior cluster quality. After calculating the Quality of Clustering (QC) values, we plot the QC against the maximum distance threshold and fit an eighth-degree polynomial to the points, then take the distance which gives the minimum value of the fitted curve as the distance threshold.

For the threshold selection process, we used the phishing data shared by Cui et al. \cite{cui2017tracking}, and selected the ones which are reported between 1 and 31 January 2016. The cluster quality and false positive rate for a corresponding selected threshold is shown in Figure~\ref{fig:threshold_selection}. As shown in the graph, we obtained the best cluster quality or minimum QC by selecting 0.251 as the distance threshold. At this point, the false positive rate is also very low, which is close to zero.

\section{Experimental Setup}
In this section, we briefly describe the experiment and methodology to evaluate our proposed method, followed by further details on the phishing and legitimate website dataset used in our study.

% \begin{figure}
%     \centering
%     \includegraphics[width=0.485\textwidth]{image/performance_comparison/ROC.png}
%     \caption{Receiver Operating Characteristic (ROC) Curve}
%     \label{fig:roc}
% \end{figure}

\subsection{Evaluation Methodology}
\color{black}
There are three experiments in this study. The first experiment aims to evaluate the performance of our prototype-based method in detecting phishing websites and compare the result to other methods, i.e., the proportional distance based method proposed by Cui et al. \cite{cui2017tracking} and the use of Doc2Vec models and the Manhattan distance as introduced by Feng et al. \cite{feng2020detection}. In this experiment, we also evaluated the performance of detection in various legitimate-to-phishing class ratios, and the detection performance when using the combination of NCD with other distance measures. In the second experiment, we simulate the incremental detection of phishing websites and assess the performance of NCD and prototype-based incremental learning. Finally, in the third experiment, we perform an analysis on the memory requirements and run-time performance for evaluating the feasibility of PhishSim implementation. \color{black}In this study, we used the LZMA algorithm when computing the NCD values at it is shown to approximate the NID better which leads to better prototype extraction and superior phishing detection performances as shown in Appendix~\ref{app:compression-algo-selection}\color{black}.

%  as described in Algorithm~\ref{alg:incremental_learning}

% also mention that the we replicated cui's method and validated our replication by testing it with their dataset.
\color{black}

% In these experiments, we focused on evaluating performance based on the following metrics:
% \begin{itemize}
%     \item Accuracy: the number of correctly classified websites divided by the total number of websites,
%     \item True positive rate (TPR): the number of correctly classified phishing websites divided by the total number of phishing websites,
%     \item False positive rate (FPR): the number of misclassified legitimate websites divided by the total number of legitimate websites.
% \end{itemize}

%\begin{table*}[]
\begin{table}
\caption{Detection Method Comparison}
\label{tab:method_comparison}
\centering
\resizebox{\textwidth}{!}{%
\begin{tabular}{@{}llll@{}}
\cmidrule(l){2-4}
 & \textbf{PhishSim} & \textbf{Cui et al. \cite{cui2017tracking}} & \textbf{Feng et al. \cite{feng2020detection}} \\ \midrule
\textbf{Website Representation} & Raw data & Tag vector (number of occurrences of specific HTML tags) & Doc2Vec vector \\
\textbf{Distance Metric} & Normalized compression distance & Proportional distance & Manhattan distance \\
\textbf{Phishing Sample Database} & Extracted phishing prototypes from prior submissions & All phishing websites from prior submissions & All phishing websites from prior submissions \\
\textbf{Clustering Algorithm} & Prototype-based clustering & Agglomerative clustering & Agglomerative clustering \\ \bottomrule
\end{tabular}%
}
%\end{table*}
\end{table}

\newpage

\subsection{Dataset}
\label{sec:dataset}
% quote the paper published in WWW
\color{black}
To evaluate the performances, we used our own dataset which is completely different from the one used in the similarity analysis (Section~\ref{sec:similarity_analysis}) and the optimal distance threshold selection process (Section~\ref{threshold_selection}). We plan to share the dataset for the experiments for future work studies.

\color{black}
To build the phishing dataset, we created a script which fetched the latest phishing URL list from PhishTank and collected phishing pages every 3 hours to maximize the number of live phishing pages. The phishing URLs that are provided by PhishTank are typically fake sign-in or login pages, or pages with input forms to steal account information. The total number of collected phishing websites are roughly 13,300 websites. To fetch the website content, we used Selenium and Chromium \cite{chromium}, instead of fetching the HTML source code using built-in \textit{wget} or \textit{curl} tools, which gave us the HTML as rendered in a browser. After performing sanitation and removing error/empty pages, we have 9,034 phishing websites which were reported by users to PhishTank \cite{phishTank} between 28 April 2020 and 22 February 2021. We also removed cloaked phishing websites, by inspecting the URL of the landing page, then removed pages whose domain is included in the top 500 domains, which indicates that there is a redirection to a legitimate page.
\color{black}

Meanwhile, to obtain a representative legitimate website dataset, we compiled legitimate pages in the Common Crawl database \cite{commonCrawl}. The legitimacy of the website is assessed by the website's popularity based on the Tranco page ranking list by Le Pochat et al. \cite{LePochat2019}. This page ranking is selected because it is claimed to be more robust against manipulation by adversaries. We use the Tranco list\footnote{Available at https://tranco-list.eu/list/7JNX.} generated on 14 March 2021, and selected the top 4,000 domains. For each domain, we collected 100 pages in Common Crawl's February/March 2021 crawl archive\footnote{Available at https://commoncrawl.s3.amazonaws.com/crawl-data/CC-MAIN-2021-10/index.html.}. After removing empty pages and error pages, we have a collection of 180,302 websites. 

\color{black}

\section{Results}
\label{sec:ncd_phishing_detection}
\color{black}
In this section, we provide further details regarding the results of the experiments. We study the performance of the phishing detection algorithm, evaluate the incremental learning framework, and analyze the memory requirements and the runtime measurement.
\color{black}

% (Algorithm~\ref{alg:ncd_classification})

\subsection{PhishSim Performance}
\label{sec:detection_performance}
\color{black}
Based on the results we obtained during the threshold selection process, we selected the maximum distance threshold value as 0.251 (Section~\ref{threshold_selection}). We evaluated the performance of our proposed phishing website detection method using prototype-based learning algorithms, with NCD as the distance metric. We applied a temporal split to the phishing dataset based on the website submission date, when allocating the data for prototype extraction (model training) and performance evaluation (testing). This is recommended in past studies over the cross-validation evaluation as cross-validations could introduce performance overestimation due to the risk of training on future data and testing on past data \cite{ho2019detecting, pendlebury2019tesseract}. We use the 7,746 phishing data submitted prior to 18 February 2021 when extracting the phishing prototypes, while the remaining 1,288 phishing data are allocated for testing. Meanwhile, we use the entire legitimate website dataset for testing. The testing dataset has a phishing to legitimate class ratio of 1 to 140. We intentionally craft the testing dataset class ratio to resemble phishing detection in the real scenarios, where typically there is one phishing page for every 100 legitimate pages \cite{Dou2017, Whittaker2010}.

We compare the performance of PhishSim with other similarity-based phishing detection approaches introduced in past studies by Cui et al. \cite{cui2017tracking} and Feng et al. \cite{feng2020detection}. Cui et al. \cite{cui2017tracking} proposed a distance metric for detecting similarity among phishing websites, the \textit{proportional distance}, and made use of hierarchical clustering to group similar phishing websites and detect whether an unknown website is clustered together with a known phishing website. Meanwhile, Feng et al. \cite{feng2020detection} proposed a method to detect phishing websites using clustering algorithms and Doc2Vec models to represent phishing and legitimate websites. To measure the website similarities, Manhattan distance is used as proposed in their study. Details on each technique, including the distance metrics and clustering algorithms, are summarized in Table~\ref{tab:method_comparison}. As the code implementation of the approaches proposed by Cui et al. \cite{cui2017tracking} and Feng et al. \cite{feng2020detection} are not publicly shared, we replicated these approaches using the same algorithms and model hyperparameters as mentioned in the research papers. 
% We also performed performance evaluations of these methods and PhishSim using a past dataset collected by Cui et al. (Appendix~\ref{appendix_past_dataset}).

% Using the dataset without URL hash-duplicates, we received similar  (Figure 28) which verifies our implementation.

% was split randomly instead of split temporally, following past studies which suggest that legitimate webpage contents are relatively stable over time \cite{Fetterly:2003:ECN:951953.952397, Xiang2011}

\begin{figure}
    \centering
    \includegraphics[width=\linewidth]{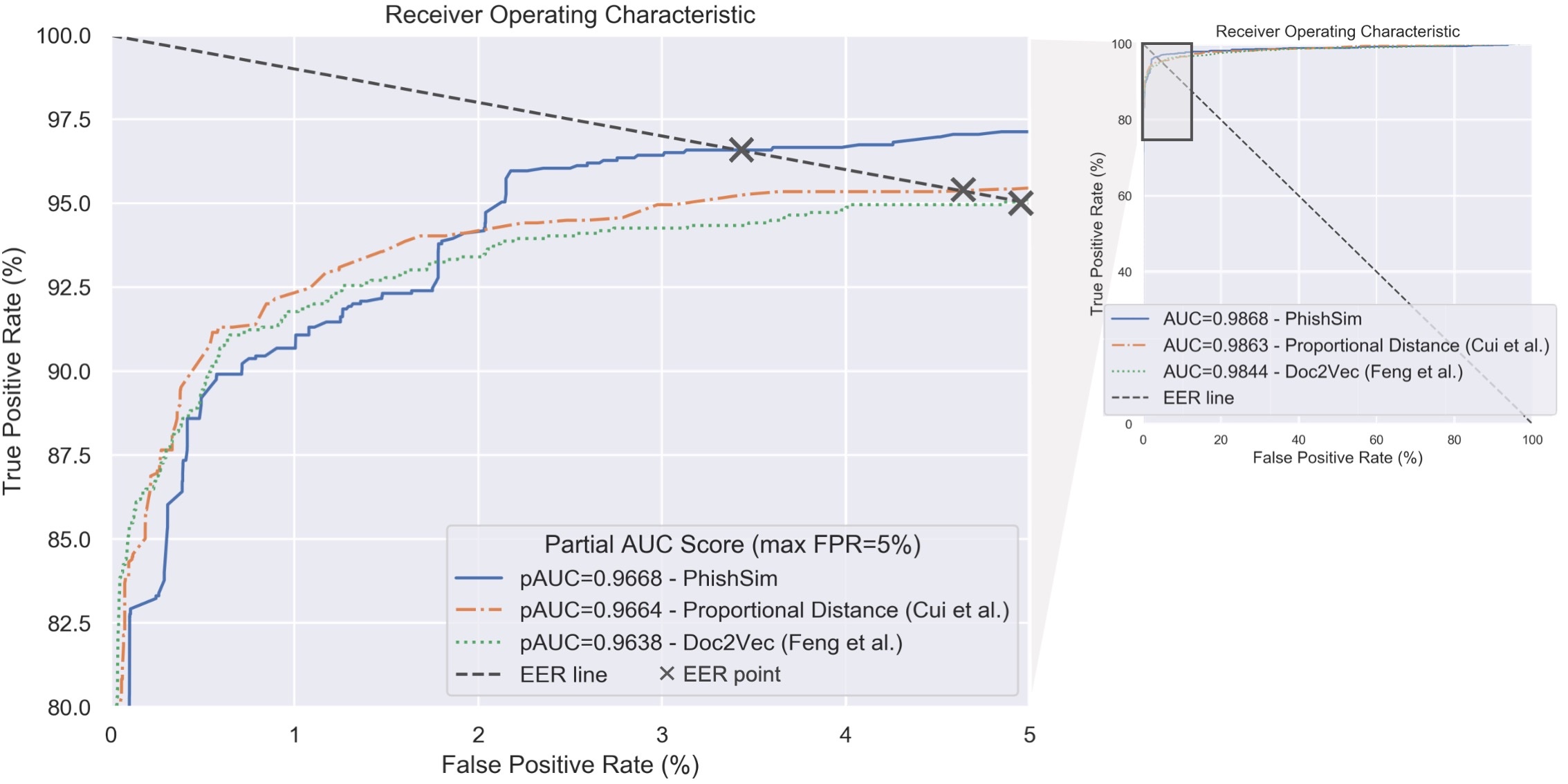}
    \caption{Receiver Operating Characteristic (ROC) Curve}
    \label{fig:roc_curve_2021}
\end{figure}

\begin{table}[]
\caption{Phishing Detection Performance Comparison (Default Distance Threshold)}
\label{tab:detection_performance}
\centering
\begin{tabular}{cccc}
%\begin{tabular}{x{1.55cm}cx{1.55cm}c}
\toprule
\textbf{Performance Metrics} & \textbf{PhishSim} & \textbf{Proportional distance \cite{cui2017tracking}} & \textbf{Doc2Vec \cite{feng2020detection}} \\ \midrule
TPR & 89.75\% & 84.08\% & 95.65\% \\
TNR & 99.42\% & 99.90\% & 93.05\% \\
FPR & 0.58\% & 0.10\% & 6.95\% \\
Accuracy & 99.36\% & 99.79\% & 93.07\% \\
G-mean & 94.47\% & 91.65\% & 94.34\% \\
\bottomrule
\end{tabular}
\end{table}

\begin{table}[]
\caption{Phishing Detection Performance Comparison (at EER Point)}
\label{tab:detection_performance_eer}
\centering
\begin{tabular}{cccc}
%\begin{tabular}{x{1.55cm}cx{1.55cm}c}
\toprule
\textbf{Performance Metrics} & \textbf{PhishSim} & \textbf{Proportional distance \cite{cui2017tracking}} & \textbf{Doc2Vec \cite{feng2020detection}} \\ \midrule
TPR	& 96.58\% & 95.34\% & 95.03\% \\
TNR & 96.59\% & 95.50\% & 95.03\% \\
FPR	& 3.41\% & 4.50\% & 4.97\% \\
Accuracy & 96.59\% & 95.50\% & 95.03\% \\
G-mean & 96.59\% & 95.42\% & 95.03\% \\
\bottomrule
\end{tabular}
\end{table}

To compare the detection performances and characteristics, we plot the receiver operating characteristic (ROC) curve based on the distance of each instance to the closest phishing sample or phishing prototype. The ROC curve is shown in Figure~\ref{fig:roc_curve_2021}. \color{black}The AUC score could be interpreted as the probability of the model assigning a phishing website with a higher similarity score than the legitimate website's similarity score. The AUC score is an aggregate measure that model's classification performance across all possible thresholds which provides a method to compare binary classifier performances regardless of the selection of distance threshold or cut-off. Based on the AUC score, PhishSim slightly outperforms other methods with an AUC score of 0.9868, indicating that it performs better in classifying phishing and legitimate websites in general. \color{black}In the case of detecting phishing attacks, we would be more interested in selecting an operating point with a low false positive rate (FPR). To observe the performance in this region, we also provide a ROC curve which focuses on the low-FPR region and the partial AUC scores of each method (maximum FPR of 5\%). As shown in Figure~\ref{fig:roc_curve_2021}, the partial AUC score also shows that PhishSim outperforms other methods with a partial AUC score of 0.9668.

% The true positive rate (TPR) of PhishSim is lower than the other approaches when FPR is lower than roughly 2\%. However, PhishSim gives higher TPR compared to other methods on the operating points where FPR is larger than 2\%.

\color{black}We also compare the detection performance when choosing the default optimal distance thresholds. As discussed in Section~\ref{sec:optimal_threshold}, we found that PhishSim works best when selecting 0.251 as the distance threshold. Meanwhile, we selected the best distance threshold of the baseline methods using the suggested value or settings as mentioned in the past study \cite{cui2017tracking, feng2020detection}. As shown in Table~\ref{tab:detection_performance}, PhishSim achieved an excellent TPR of almost 90\% with a low FPR of 0.58\%. While Doc2Vec achieves a higher TPR, the FPR at this operating point is also significantly higher, which would not be ideal for phishing detection. We also measured the G-mean, or the geometric mean of the TPR and true negative rate (TNR), which provides a better way to evaluate a model in a highly imbalanced classification problem, which is similar to the ratio between phishing to legitimate websites in real scenarios. As shown in Table~\ref{tab:detection_performance}, PhishSim outperformed other methods with a G-mean of 94.47\%, indicating a better overall performance in detecting phishing with the real imbalanced class distribution. Furthermore, we also compared the performance of each method when selecting the EER point as the operating point, which indicates the operating point when the false negative rate and the false positive rate have the same value. As shown in Table~\ref{tab:detection_performance_eer}, PhishSim has superior performance compared to other methods, by achieving TPR, TNR, and accuracy, with the lowest FPR. It is also shown that PhishSim gives the best G-mean score, which shows the ability to provide the best compromise between achieving a high TPR while still having a high TNR.
\color{black}

% To observe whether the class ratio of the dataset affects the detection performance, we also performed an experiment to observe the false positive rate in various phishing to legitimate ratios. We randomly sample the legitimate website data to achieve a phishing to legitimate ratio of 1:1 to 1:100. The results are shown in Figure~\ref{fig:fpr_varied_class_ratio}. While slightly fluctuating, it is shown that each method was able to maintain the false positive rate in a certain region regardless of the dataset class ratio.

% As we are using the whole 1,288 phishing data in the testing dataset, the TPR would remain constant in this experiment, and would be identical to the TPR values in Table~\ref{tab:detection_performance}. 

% \begin{figure}
%     \centering
%     \includegraphics[width=\linewidth]{image/new_results_unbalanced_dataset/FPR_varied_class_ratio.png}
%     \caption{False Positive Rate in Various Class Ratios}
%     \label{fig:fpr_varied_class_ratio}
% \end{figure}

\begin{figure}
    \centering
    \includegraphics[width=0.8\linewidth]{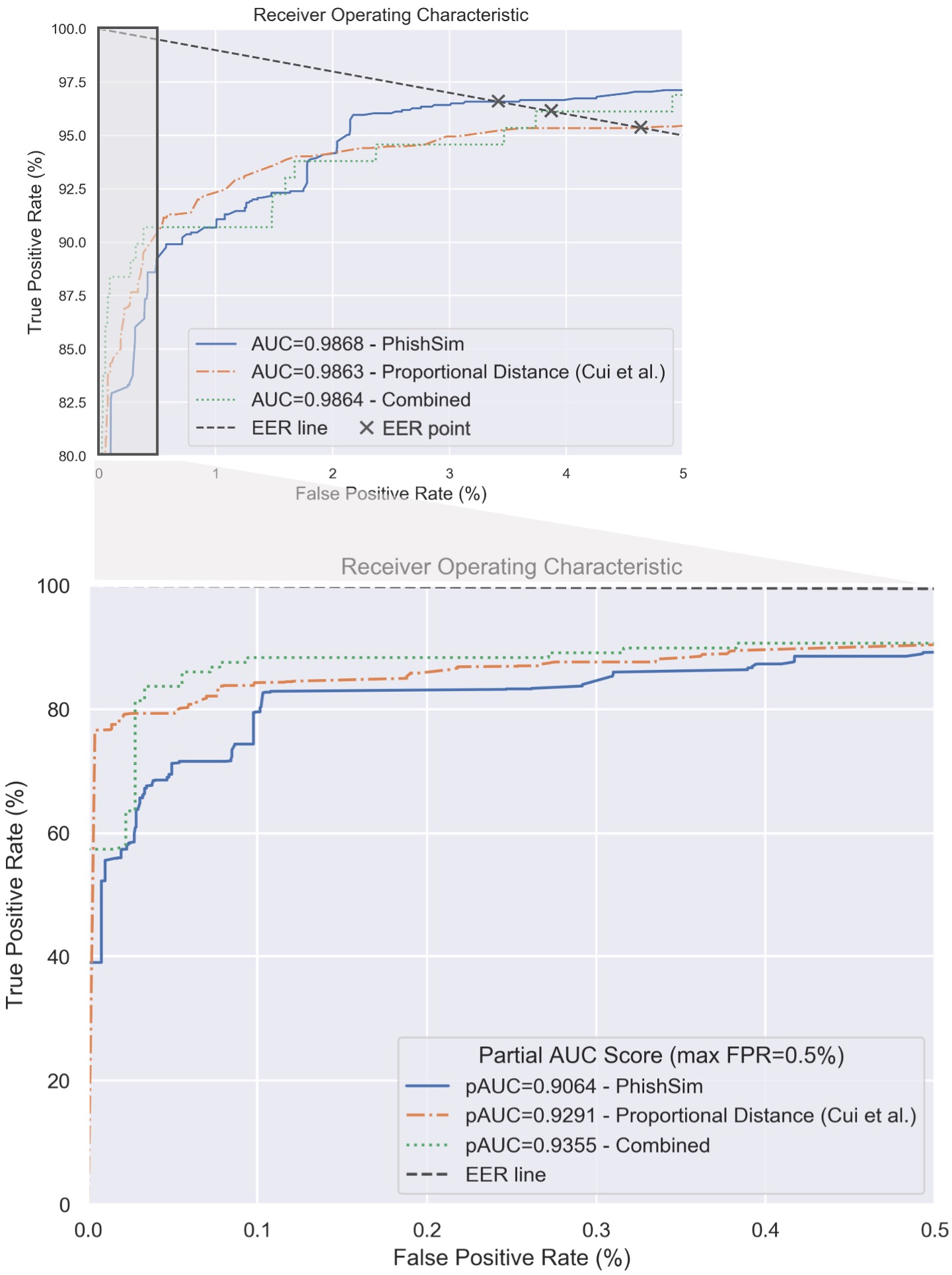}
    \caption{Combining NCD and Proportional Distance}
    \label{fig:roc_curve_lr}
\end{figure}

\begin{table}[]
\caption{Performance of PhishSim with Proportional Distance (FPR$<$0.1\%)}
\label{tab:combined_performance}
\centering
\begin{tabular}{cccc}
%\begin{tabular}{x{1.55cm}cx{1.55cm}c}
\toprule
\textbf{Performance Metrics} & \textbf{PhishSim} & \textbf{Proportional distance \cite{cui2017tracking}} & \textbf{Combined} \\ \midrule
TPR & 79.58\% & 84.32\% & \textbf{88.37\%} \\
TNR & 99.90\% & 99.90\% & \textbf{99.91\%} \\
FPR & 0.10\% & 0.10\% & \textbf{0.09\%} \\
Accuracy & 99.76\% & 99.79\% & \textbf{99.82\%} \\
G-mean & 89.16\% & 91.78\% & \textbf{93.96}\% \\
\bottomrule
\end{tabular}
\end{table}

\vspace{0.35cm}
\subsection{PhishSim for Improving Existing Methods}
\color{black}
We also observed if the use of PhishSim in combination with other methods would improve the phishing detection performance. In this experiment, we combined the two best performing methods in terms of AUC score, which are PhishSim and the proportional distance based method by Cui et al. \cite{cui2017tracking} and trained a logistic regression model to perform a weighted sum on the NCD and proportional distance values.

For this experiment, we used the dataset that we collected, as mentioned in Section~\ref{sec:dataset}. We allocated 90\% of the phishing and legitimate website data for training the logistic regression model and used the remaining 10\% of the phishing and legitimate website data for testing the model's performance, while keeping the phishing to legitimate class ratio to 1:140 in both training and testing dataset. To improve the trained logistic regression model's performance, we applied the SMOTE oversampling method on the training dataset to generate synthetic phishing data to provide more samples.

The performance of the logistic regression model is shown in Figure~\ref{fig:roc_curve_lr}. The AUC score shows a similar performance in general. However, in a very low FPR region (maximum FPR of 0.5\%), the combined method was able to achieve a higher partial AUC score compared to PhishSim and proportional distance alone. To compare the performance, we select an operating point where the FPR is around 0.1\% which is considered an acceptable false positive rate for phishing detection in real scenarios, similar to some past studies \cite{cui2017tracking, Whittaker2010}. The TPR, FPR, and accuracy of the combined method at this operating point is shown in Table~\ref{tab:combined_performance}. It is shown that combining these methods resulted in a great improvement in terms of TPR, with a 5\% difference from using proportional distance alone, while having a slightly lower FPR of 0.09\% and a higher accuracy of 99.82\%.
\color{black}

\color{black}

\subsection{Incremental Learning}
\label{sec:inc_detection_performance}

% (Algorithm~\ref{alg:incremental_learning})

\color{black}
With the selected threshold (Section~\ref{threshold_selection}), we also evaluated the performance using the incremental learning method. Similar to the previous experimentation, we also applied a temporal split to the dataset. The phishing dataset was sorted and divided by its week of submission for each iteration of the incremental learning process. In each iteration, we treated the last week of data as the \emph{testing data} and performed classifications to evaluate the detection performance using the prototypes extracted in earlier months. Following this, we performed prototype extraction on the testing data at each iteration to learn new phishing prototypes, simulating the process of weekly detection and database updates of anti-phishing systems. Using the testing data from 28 May 2020 to 22 February 2021, we performed evaluations for roughly 39 weeks. As a comparison, we also implemented the hierarchical clustering method proposed by Cui et al. \cite{cui2017tracking} and Doc2Vec based method by Feng et al. \cite{feng2020detection} in an incremental learning experimental setup. Note that instead of updating the clustering model, we performed clustering from scratch in every iteration when implementing the method by Cui et al. \cite{cui2017tracking} and Feng et al. \cite{feng2020detection}, since we are unable to find an approach to perform the clustering incrementally using this method. The performance of PhishSim compared to these methods at the last iteration of the incremental learning experiment is shown in Table~\ref{tab:inc_detection_performance}, which is consistent with the results from the non-incremental learning setting in Table~\ref{tab:detection_performance}. \color{black}We also included the ROC curve and performance at the EER point in Table~\ref{tab:inc_detection_performance_eer}.\color{black}

\begin{table}[]
\caption{Phishing Detection Performance Comparison for Incremental Learning (Default Distance Threshold)}
\label{tab:inc_detection_performance}
\centering
\begin{tabular}{cccc}
%\begin{tabular}{x{1.55cm}cx{1.55cm}c}
\toprule
\textbf{Performance Metrics} & \textbf{PhishSim} & \textbf{Proportional distance \cite{cui2017tracking}} & \textbf{Doc2Vec \cite{feng2020detection}} \\ \midrule
TPR & 89.52\% & 84.08\% & 94.25\% \\
TNR & 99.22\% & 99.91\% & 95.74\% \\
FPR & 0.78\% & 0.09\% & 4.26\% \\
Accuracy & 97.11\% & 96.47\% & 95.42\% \\
G-mean & 94.25\% & 91.66\% & 95.00\% \\ \bottomrule
% AUC Score & 98.65\% & 98.64\% & 98.64\% \\ \bottomrule
\end{tabular}
\end{table}

\begin{figure}
    \centering
    \includegraphics[width=0.6\linewidth]{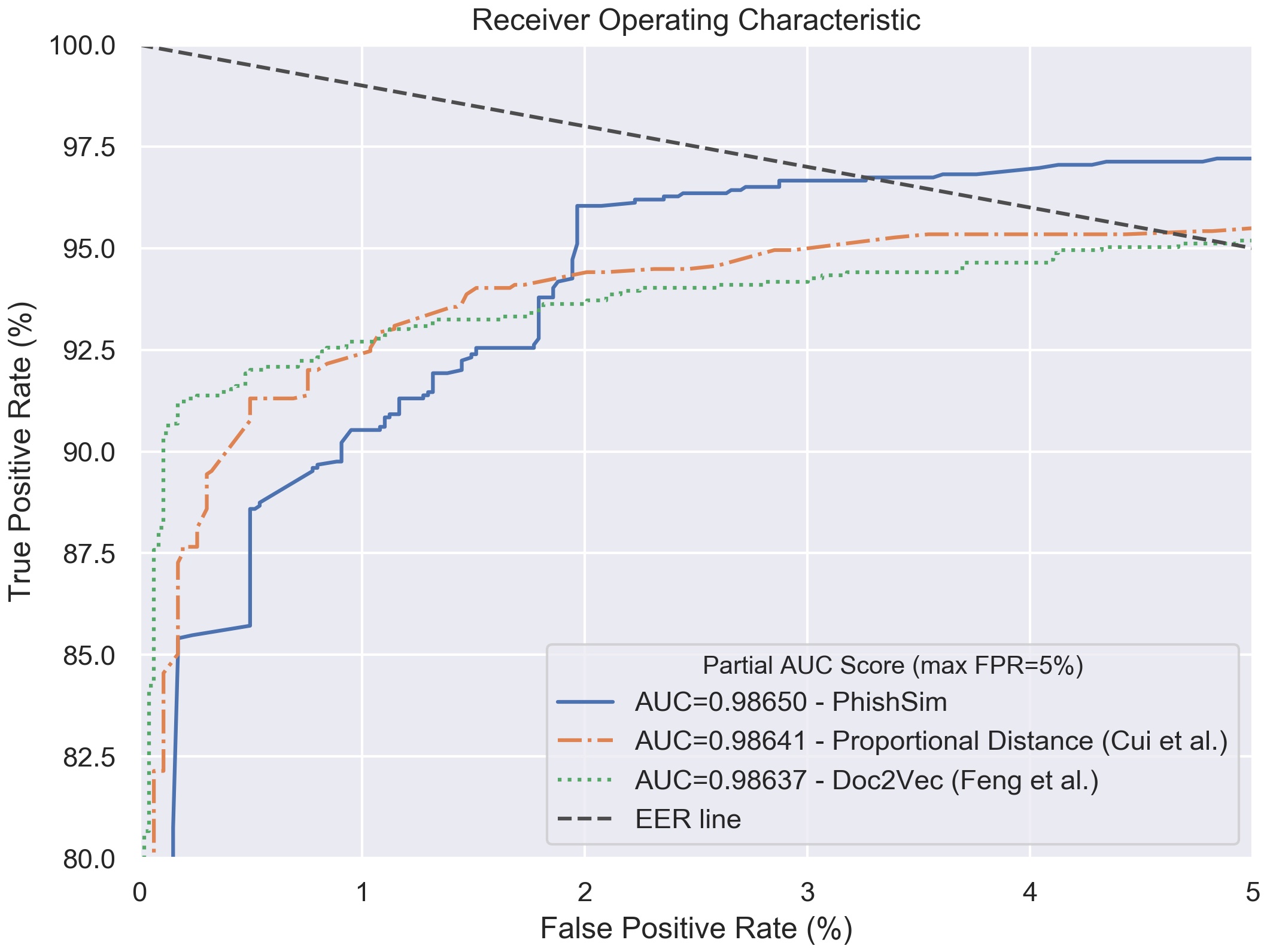}
    \caption{Receiver Operating Characteristic (ROC) Curve (in Incremental Learning Experiment)}
    \label{fig:inc_roc_curve_2021}
\end{figure}

\begin{table}[]
\caption{Phishing Detection Performance Comparison for Incremental Learning (at EER Point)}
\label{tab:inc_detection_performance_eer}
\centering
%\begin{tabular}{x{1.55cm}cx{1.55cm}c}
\begin{tabular}{cccc}
\toprule
\textbf{Performance Metrics} & \textbf{PhishSim} & \textbf{Proportional distance \cite{cui2017tracking}} & \textbf{Doc2Vec \cite{feng2020detection}} \\ \midrule
TPR & 96.66\% & 95.34\% & 95.11\% \\
TNR & 96.74\% & 95.57\% & 95.10\% \\
FPR & 3.26\% & 4.43\% & 4.90\% \\
Accuracy & 96.72\% & 95.52\% & 95.10\% \\
G-mean & 96.70\% & 95.46\% & 95.10\% \\
\bottomrule
\end{tabular}
\end{table}

\begin{figure}
    \centering
    \includegraphics[width=0.8\linewidth]{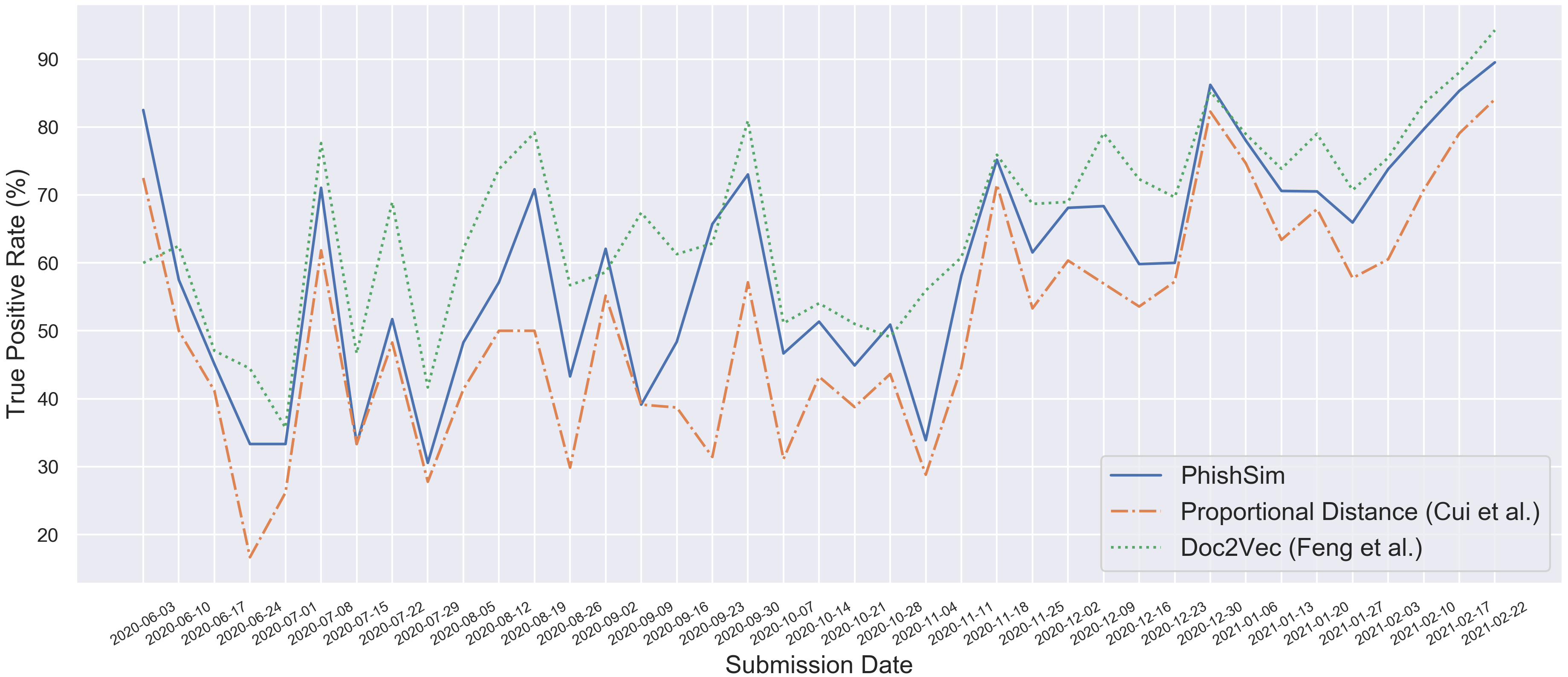}
    \caption{TPR in Incremental Learning Setting}
    \label{fig:inc_tpr}
\end{figure}

\begin{figure}
    \centering
    \includegraphics[width=0.8\linewidth]{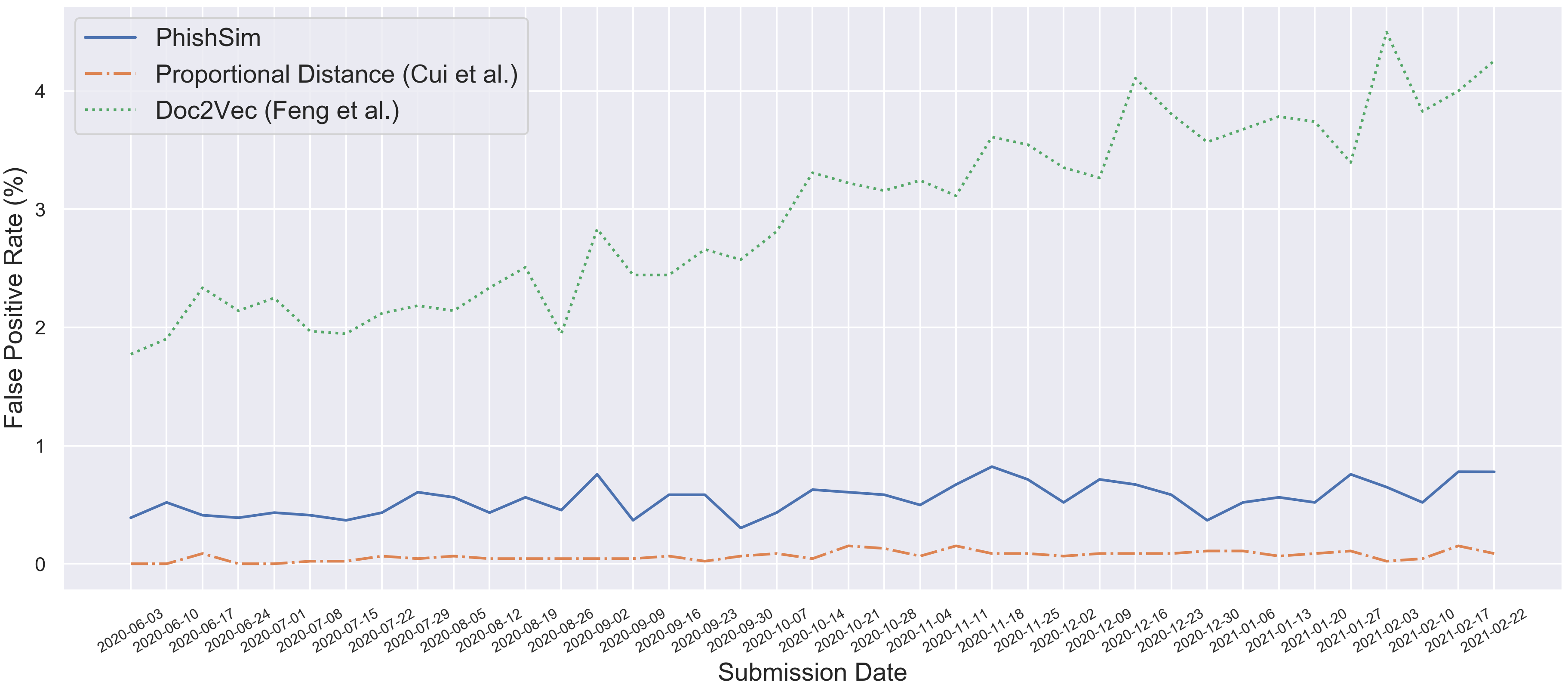}
    \caption{FPR in Incremental Learning Setting}
    \label{fig:inc_fpr}
\end{figure}

Furthermore, we also observed the true positive rate and false positive rate at each iteration, which are provided in Figure~\ref{fig:inc_tpr} and Figure~\ref{fig:inc_fpr}. As shown in Figure~\ref{fig:inc_tpr}, the Doc2Vec based method by Feng et al. \cite{feng2020detection} outperforms the other methods in most iterations. However, it is also shown in Figure~\ref{fig:inc_fpr} that the false positive rate of the Doc2Vec-based method is relatively high and seems to increase over time, which would not be suitable for phishing detection. On the other hand, PhishSim consistently outperforms the proportional distance based method by Cui et al. \cite{cui2017tracking} in terms of true positive rate with a TPR of nearly 90\%. While having a high TPR, PhishSim maintains a low and stable FPR of under 0.8\% in each iteration.

% \begin{figure*}
%     \centering
%     \includegraphics[width=0.925\textwidth]{image/new_results_unbalanced_dataset/Incremental_accuracy.png}
%     \caption{Accuracy in Incremental Learning Setting}
%     \label{fig:inc_acc}
% \end{figure*}

\subsection{Memory Requirements and Run-time Analysis}
\color{black}
Besides evaluating the detection performance, we also analyzed the memory requirements and run-time performance of PhishSim to evaluate the feasibility of deployment in real scenarios. To observe the memory requirements, we analyze the number of prototypes extracted at each iteration in the incremental learning experiment in comparison to the number of phishing websites these prototypes represent. \color{black}We also computed the amount of time needed to process every website using a 6-core 3.70 GHz Intel(R) Xeon(R) W-2135 CPU.\color{black}

\begin{figure}
    \centering
    \includegraphics[width=0.6\linewidth]{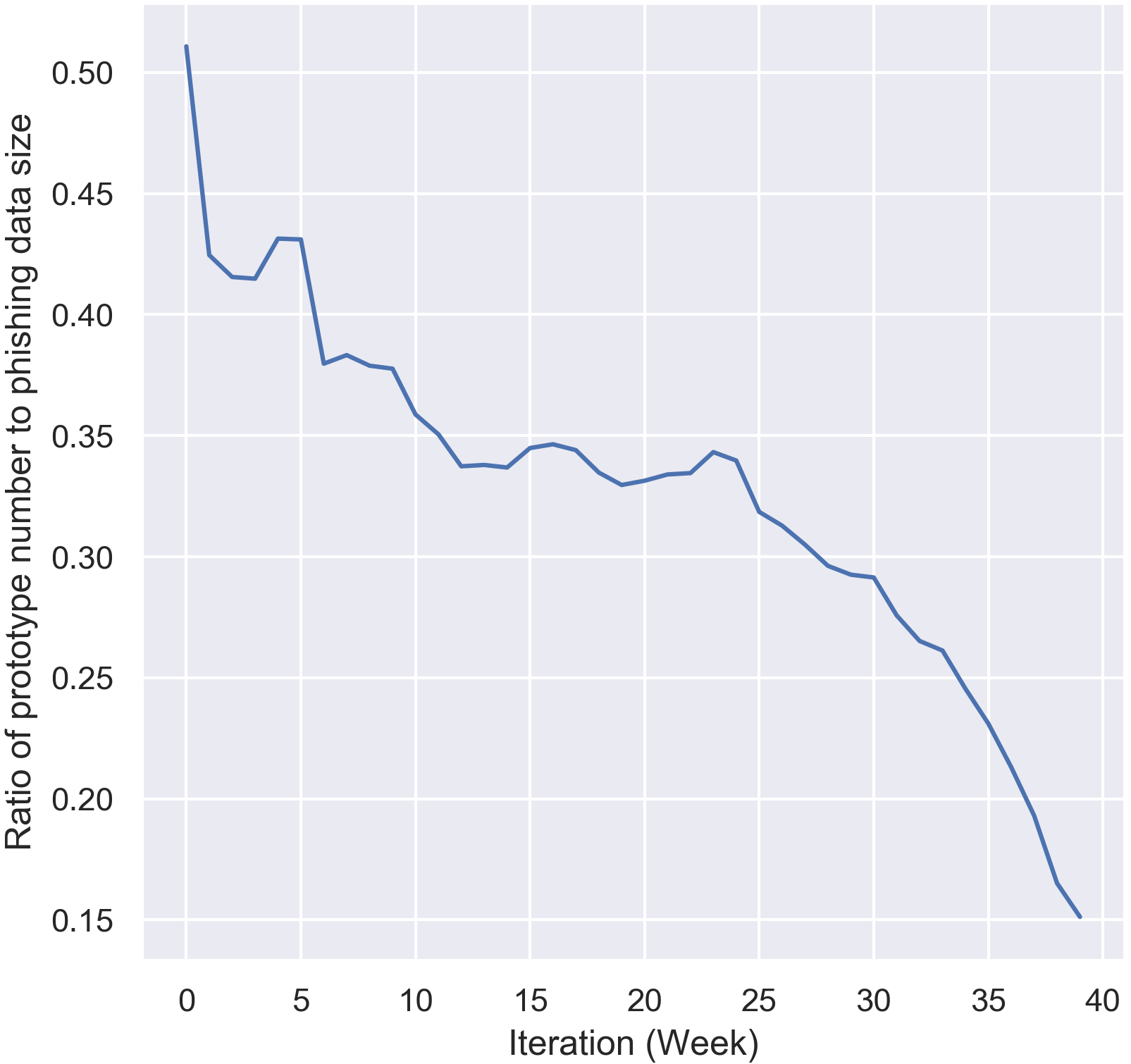}
    \caption{Ratio of prototype number to phishing data size}
    \label{fig:num_of_prototypes}
\end{figure}

Figure~\ref{fig:num_of_prototypes} depicts the compression ratios at each iteration in the incremental learning experiment (as discussed in Section~\ref{sec:inc_detection_performance}). Compression ratio in this context is defined as the ratio between the number of prototypes extracted relative to the total number of phishing websites they represent. As the detection system incrementally learns, the number of prototypes increases as new website samples are learned. However, compression ratio gradually decreases over time, showing that the system was able to extract meaningful phishing prototype representations. At the last iteration, around 1,366 prototypes were extracted which represents 9,034 phishing websites, giving a compression ratio of 0.15. The phishing HTML DOM data size average is around 2.02 KB with a standard deviation of 3.81 KB. A larger standard deviation relative to the mean would indicate that the data distribution is heavily right-skewed with a number of data outliers. Therefore, we use the median of the phishing website HTML DOM size in this analysis to predict the data storage consumption. Based on our experiment, the number of prototypes is 1,366 at the last iteration. With a median of around 727 B, storing 1,366 phishing prototypes would take roughly 0.947 MB of data storage.

To evaluate the run-time performance, we analyzed the time needed to perform classification on each website. To decide whether a website is legitimate or phishing, we perform pairwise NCD calculation between the website and each phishing prototype. The length of the compressed phishing prototype can be precomputed before classification, while the length of the compressed website only needs to be computed once. Thus, the process which would take time the most is compressing the concatenation of the website and each phishing prototype, which cannot be precomputed and should be performed individually for each prototype. The total amount of time to process a website would depend on the number of phishing prototypes. \color{black}With the assumption that the length of each compressed prototype file is precomputed and the number of prototypes in the database is 1,366, we found that it would take around 0.3 seconds to process a single website.\color{black}

\begin{figure*}
    \centering
    \includegraphics[width=\linewidth]{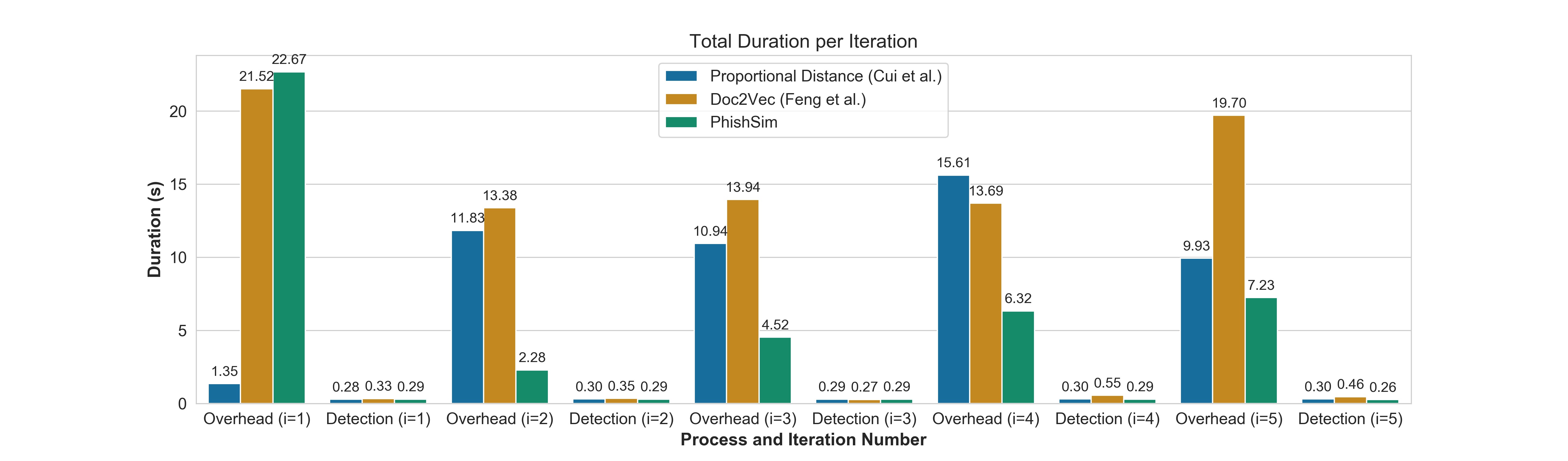}
    \caption{Total Process Duration at $1^{st}$ to $5^{th}$ Iteration}
    \label{fig:exec_time}
\end{figure*}

\color{black}In the incremental learning experiment (Section \ref{sec:inc_detection_performance}), we also measured the total amount of time taken by PhishSim and other methods in past studies to perform detection and any other processes to prepare the detection model, which include prototype extraction (PhishSim), HTML tag computation and distance matrix construction (Proportional Distance \cite{cui2017tracking}), Doc2Vec model training, vector inference, and distance matrix construction (Doc2Vec \cite{feng2020detection}). The total process duration at the first five iteration is shown in Figure~\ref{fig:exec_time}. As shown in Figure~\ref{fig:exec_time}, PhishSim takes less than 0.3 seconds to perform detection, which is relatively similar to the time taken by the other methods. While PhishSim has the highest overhead on the first iteration, it takes the least amount of time to update the model on the next iterations compared to other methods, indicating the PhishSim's detection model can be updated incrementally faster.\color{black}

% While the empirical results show that our proposed method takes longer time to process a single website, the time-complexity analysis show that the detection system would scale better with PhishSim. Performing classification in each iteration would take $O(kN)$ time, where $k$ is the number of prototypes and $N$ is the data size. As shown in Figure~\ref{fig:num_of_prototypes}, the prototype extraction algorithm was able to select meaningful prototypes with a compression ratio of 0.18 at the end of the iteration. In contrast, the hierarchical clustering algorithm in \cite{cui2017tracking} uses a greedy algorithm which takes a complexity of $O(N)^2$, indicating that the time to perform classification will grow quadratically as the number of phishing websites increases.

%To compare with the state-of-the-art detection method by Cui et al. \cite{cui2017tracking}, we also performed a time-complexity analysis. Let $n$ and $N$ be the size of the training and test datasets respectively, and let $k$ be the number of prototypes. Algorithm~\ref{alg:prototype_extraction} takes $O(kn)$ time and Algorithm~\ref{alg:ncd_classification} runs in $O(kN)$ time. Therefore, each iteration of Algorithm~\ref{alg:incremental_learning} takes $O(k(n+N))$ time. In contrast, the hierarchical clustering algorithm in \cite{cui2017tracking} uses a greedy algorithm which takes a complexity of $O(n+N)^2$, indicating that our detection method is more efficient.

\color{black}

\section{Discussion}
\label{sec:discussion}

\subsection{Detection Method and Performance}
This study proposes a feature-free method for detecting phishing websites, which we argue would be suitable with the dynamic characteristics of phishing attacks. As reported by PhishLabs \cite{phishlabs2019}, phishing attackers have been persistently adapting to new opportunities and changing the manner in which they adapt. Furthermore, a study by Cui et al. \cite{cui2017tracking} found that around 90\% of phishing websites are variations or replicas of other phishing websites. There has also been an increased use of phishing kits, which ease the process of setting up phishing websites, enabling attackers to launch a large number of phishing attacks in a short period of time. As discussed in Section~\ref{sec:brand_specific_analysis}, we found that these kits make the process of changing the phishing website layout and style more effortless.

Based on the experiments in Section~\ref{sec:detection_performance} and Section~\ref{sec:inc_detection_performance}, our proposed method is able to outperform the hierarchical-clustering based method by Cui et al. \cite{cui2017tracking} and Doc2Vec based method by Feng et al. \cite{feng2020detection} based on its AUC score. PhishSim achieves a high TPR of around 90\% while maintaining a low FPR of 0.58\%. We can also apply this method to build a model which can incrementally learn using one week of data to perform phishing detection with relative success. Evaluating our detection method in comparison with state-of-the-art methods, it has also been shown that PhishSim is able to outperform past studies with significant improvements in terms of true positive rate and with a relatively low FPR which is suitable for phishing detection. It has also been shown that we are able to improve the phishing detection performance when combining PhishSim with proportional distance, as shown in Table~\ref{tab:combined_performance}. Using this combined method, we could operate at an even lower FPR of 0.09\% with a TPR of 88.37\% and accuracy of 99.82\%.

% Furthermore, performance evaluation on the past dataset shared by Cui et al. also indicated that our proposed method is significantly better in detecting phishing websites as indicated by the true positive rate and AUC score (Appendix~\ref{appendix_past_dataset}).

The method that we proposed is a feature-free method, which means that our method is not tied to a fixed website representation (Table~\ref{tab:method_comparison}) and is more robust compared to feature-based methods which rely on specific features. On the other hand, the method by Cui et al. \cite{cui2017tracking} uses the proportional distance which measures website similarity based on the number of occurrences of a set of predefined HTML tags and assumes that similar websites have a nearly identical number of HTML tag counts. \color{black}While phishing websites that use custom HTML tags are not very common currently, the use of WebComponents \cite{mozillaWebComponent} provide a method to perform novel website obfuscation technique using custom HTML tags to avoid detection from content-based methods which heavily relies on the analysis of the HTML content. While adding HTML tag definition and model training are possible, phishing attacks are very dynamic and aggressive with changes in evasive phishing campaign methods every 37 days in average \cite{microsoftEvasivePhishingCampaign}. Having the flexibility to perform detection based on similarities without reliance on a preset corpus would be a great benefit to counteract the dynamics of phishing. \color{black}Similarly, the approach by Feng et al. \cite{feng2020detection} uses a trained Doc2Vec model to generate vector representation for each website. It is not guaranteed that the generated vectors are always representative over time. To have an accurate detection, these methods rely on the assumption that the website representation would always remain relevant, which might not always be the case. Some features that were previously used to detect phishing websites accurately (e.g., bad forms, bad action fields, and non-matching URLs \cite{Xiang2011}) may no longer be relevant to phishing attacks. A past study investigated the relevance of these features in a phishing dataset collected in 2020 and has shown that only 8.90\% of the phishing websites contain bad forms, while only 19.95\% and 47.92\% of them contain bad action fields and non-matching URLs respectively \cite{purwanto2020phishzip}.

On the other hand, NCD as a similarity measure is universal and is purely based on the shared information between measured websites. Every phishing cluster is represented by a data instance, while the NCD value is used for measuring the relationship between all pairs of websites. Furthermore, the use of the Furthest Point First (FPF) algorithm provides a systematic way for prototype learning such that it does not require the data to be represented in a fixed vector representation or data structure. This is unlike most clustering and classification algorithms which require the data to be represented as a vector of representative feature values. Once the selected feature values are no longer relevant for phishing attacks, there is a risk of missing malicious phishing attacks. Meanwhile, our feature-free method would still be able to continuously identify these attacks with a maintained up-to-date phishing data and without the reliance on a specific feature set.

Robustness against code obfuscation is generally a concern when the detection approach evaluates the website’s HTML source code. However, as stated in Section~\ref{sec:system-overview}, PhishSim evaluates the HTML of the rendered page instead of the HTML source code. A web browser will produce the same HTML code regardless of the HTML obfuscation or hidden alterations in the CSS or JavaScript files. Furthermore, we also removed the text and HTML comments in the content prior to performing compression, leaving only the HTML tags which are rendered and visually shown on the browser. Thus, the performance will also be unaffected by addition of invisible elements in the HTML.

\subsection{Limitations and Future Work}
There are some limitations to our proposed method. First, it may not be able to detect zero-day attacks or new variants of phishing websites, as our method is primarily focused on detecting variations of known attacks. However, we argue that our proposed method would improve the quality of detection done by experts, by using various machine learning algorithms, and by providing a framework to continuously learn and adapt to new cases of phishing, without the need to choose a fixed representation of phishing instances.
% Our method could preserve knowledge by retaining the phishing data as it is.
% can be possibly combined with ML based methods to obtain collection of phishing websites

Moreover, there are some challenges in the practical implementation of this method. It is possible that a certain phishing prototype becomes obsolete which then makes the prototype irrelevant for detection. Therefore, ideally, there needs to be a way to effectively maintain the prototype collection so that only the useful prototypes are retained in the database. This is, however, currently beyond the scope of our paper. The proposed method can also be improved in the future by designing methods to effectively maintain useful prototypes, as well as by removing instances which are no longer relevant. The ability to remove irrelevant prototypes would be beneficial to reduce the storage size of the prototype database.

Our proposed phishing detection approach uses a prototype-based clustering and classification approach. This helps us to make the algorithm's decisions more interpretable, as it clarifies what each cluster or class corresponds to, by observing the prototypes. For future work, it would be interesting to further investigate the relationship between the set of prototypes that the algorithm finds and the phishing dataset, and how they evolve over time.

\color{black}
Furthermore, the study of PhishSim was mainly motivated by a previous work \cite{cui2017tracking}, which found that most of the phishing websites are replicas with similar HTML contents. Currently, performing a comprehensive comparison of using content similarity over visual similarity would be a challenging task, due to the unavailability of a large-scale high-quality dataset that contains both the rendered page's HTML and its screenshot. Future studies that focus on a comprehensive study of the content-based and visual-based methods using a standard website dataset would be beneficial in improving phishing detection studies.
\color{black}

\section{Conclusions}
\label{sec:conclusions}

In this paper, we propose a feature-free method for detecting phishing websites using the Normalized Compression Distance (NCD) which measures website similarity by compressing them, eliminating the need to perform any feature extraction nor any reliance on a specific set of website features. This method examines the HTML source codes of webpages and computes their similarity with known phishing websites. We propose the use of the Furthest Point First algorithm to perform phishing prototype extractions, in order to select instances that are representative of a cluster of phishing webpages. We also introduce the use of an incremental learning algorithm as a framework for continuous and adaptive detection without re-performing new feature extraction when concept drift occurs. Evaluating the performance on a recent large-sized dataset, our proposed method is shown to outperform past studies in detecting phishing websites with an AUC score of 98.68\%, a high true positive rate of around 90\% while maintaining a low FPR of 0.58\%.

\section*{Acknowledgment}
This work has been supported by the Cyber Security Cooperative Research Centre Limited, whose activities are partially funded by the Australian Government's Cooperative Research Centres Programme.

Rizka Widyarini Purwanto was supported by a UNSW University International Postgraduate Award (UIPA) scholarship. Any opinions, findings, and conclusions or recommendations expressed in this paper are those of the authors and do not necessarily reflect the views of the scholarship provider.

%\ifCLASSOPTIONcaptionsoff
%  \newpage
%\fi



%\bibliographystyle{IEEEtran}
\bibliographystyle{IEEEtranS}
\bibliography{PhishSim}

\appendix

\section{\color{black}Normalized Compression Distance\color{black}}
\label{appendix_ncd}

Normalized Compression Distance (NCD) is an application independent  information theoretic  method for measuring the similarities between two objects. It is a parameter-free similarity measure which uses compression algorithms to perform data clustering and classification in a broad range of applications. With this method, useful knowledge can be obtained from data without prior domain expertise as it operates on generic file objects, regardless of their format, structure, or semantics \cite{li2004similarity}.

NCD computes the distance between two files by observing the result of compressing both files together and comparing this with the result of compressing each file separately. The key concept behind NCD is that two very similar files would compress much more effectively when combined prior to compression, compared with the total file size when compressed separately. On the other hand, compressing two files with little or nothing in common would not be as beneficial as compressing them separately \cite{cilibrasi2005clustering}.

NCD is motivated by the idea that the similarity of two objects can be measured by how easy it is to transform one object into the other. This concept is formally expressed by the information distance $E(x,y)$, which is defined as the length of the shortest binary program to compute $y$ from $x$ or $x$ from $y$, and can be rewritten as
\begin{equation}
\label{eq:info_dist}
    E(x,y) = max\{K(x|y), K(y|x)\}
\end{equation}
The information distance is based on the notion of \emph{Kolmogorov complexity}, $K(x)$, which depicts the length of the shortest program that computes $x$ \cite{cilibrasi2005clustering}.

The normalized version of $E(x,y)$, called the normalized information distance (NID), is defined as
\begin{equation}
\label{eq:nid}
    NID(x,y) = \frac{max\{K(x|y),K(y|x)\}}{max\{K(x),K(y)\}}
\end{equation}
NID itself satisfies the conditions of a metric and can represent the similarity between two arbitrary entities according to the dominating shared features between both objects \cite{li2004similarity}. Unfortunately, the NID is based on the Kolmogorov complexity, which is not computable. This indicates that NID cannot be used directly. However, Cilibrasi and Vitanyi \cite{cilibrasi2005clustering} demonstrate that we can approximate the Kolmogorov complexity, and introduce the notion of NCD as a practical similarity metric approximating NID using a real-world compression algorithms $C$, which is defined as,
\begin{equation}
\label{eq:ncd}
    NCD(x,y)=\frac{C(xy)-min\{C(x),C(y)\}}{max\{C(x),C(y)\}}
\end{equation}

The NCD is a non-negative number between 0 and $1+\epsilon$ which depicts how similar two objects are. Highly similar files would give smaller NCD values, while distinctive files would give NCD values closer to 1. Meanwhile, the $\epsilon$ in the upper bound represents a small error caused by imperfections in the compression algorithms which is usually a value below 0.1 for most standard algorithms \cite{cilibrasi2005clustering}.

The approximation of the denominator of NID in Equation \ref{eq:nid} given a compressor $C$ is quite obvious, giving us the denominator of NCD in Equation~\ref{eq:ncd}. However, the approximation of the NID numerator is not as straightforward. The numerator in Equation~\ref{eq:nid} can be expressed as
\begin{equation}
\label{eq:nid_numerator}
    max\{K(x,y)-K(x),K(x,y)-K(y)\}
\end{equation}
with
\begin{equation}
\label{eq:kolmogorov_concat}
    K(x,y)=K(xy)=K(yx)
\end{equation}
where $xy$ or $yx$ depicts the concatenation of $x$ and $y$, and $K(x,y)$ denotes the length of the shortest program to compute $(x,y)$. The approximation is expressed using $K(x,y)$ since in practice, it would be easier to perform compression on the concatenation $xy$. Equation~\ref{eq:nid_numerator} can be best approximated by
\begin{equation}
\label{eq:nid_numerator_approximation}
    min\{C(xy),C(yx)\}-min\{C(x),C(y)\}
\end{equation}
In our experiment, $C(x,y)$ is used instead of $min\{C(xy),C(yx)\}$ as proposed in \cite{cilibrasi2005clustering}. Here, we assume that $C$ is \emph{symmetric}, i.e., $C(xy) = C(yx)$. This is justified by previous experiments \cite{cilibrasi2005clustering} using various block-coding based and stream-based compression algorithms, where it has been shown that the results are symmetric or only produce small deviations from symmetry.

\color{black}
\section{Compression Algorithm Selection}
\label{app:compression-algo-selection}
% Based on the result in our ACM CCS 2021 paper draft version
% (jupyter notebooks are in Project 30)
In this section, we performed an evaluation on the performance of PhishSim using various compression algorithms (i.e. zlib, bz2, LZMA, and gzip). For this comparison, we used a different set of phishing and legitimate website data from the datasets used in our main experiment. To build the phishing dataset for this evaluation, we use the list of 9,245 phishing website URLs reported by user to PhishTank \cite{phishTank} between 7 November 2008 and 28 March 2020. Meanwhile, to obtain representative legitimate website dataset, we collected URLs of non-phishing pages which are associated with the brands or organizations that are commonly targeted by phishing attacks, especially its login page. This data collection process makes use of the Google Custom Search API \cite{googleCse} to perform custom search using a specific query and specifying the search results to be pages from a specific domain. To collect the legitimate webpage URLs, the phishing target list by PhishTank and Alexa top 1 million sites \cite{alexa} were used. Each domain in the list was searched using the Google Custom Search Engine API and we collected the login page as well as the top 100 URLs the search. In total, our legitimate dataset consists of 10,869 legitimate websites, which were collected between February and April 2020. We applied a temporal split to the phishing dataset based on the website submission date, when allocating the data for prototype extraction (model training) and performance evaluation (testing). The training dataset consists of 8,511 phishing websites and 8,511 legitimate websites, while the testing dataset comprises 734 phishing and 734 legitimate websites.

The phishing detection performance comparison results are provided in Table~\ref{tab:compression_comparison}. As shown in this table, the use of zlib algorithm leads to the best performance in terms of TNR and AUC score. However, the LZMA algorithm achieved a significantly higher TPR of 74.114\% which is 13\% higher than the TPR of zlib algorithm, indicating a superior ability in classifying the phishing websites. Furthermore, LZMA also outperformed other compression algorithms in terms of the accuracy and G-mean score, which computes the geometric mean of TPR (sensitivity) and TNR (specificity).

\begin{table}[]
\caption{Performance Comparison of Various Compression Algorithms}
\label{tab:compression_comparison}
\centering
\begin{tabular}{ccccc}
%\begin{tabular}{x{1.5cm}ccccc}
\toprule
Performance Metrics & zlib & bz2 & LZMA & gzip \\
\midrule
TPR & 61.717\% & 66.757\% & \textbf{74.114\%} & 66.076\% \\
TNR & \textbf{96.594\%} & 96.322\% & 94.142\% & 94.550\% \\
Accuracy & 79.155\% & 81.540\% & \textbf{84.128\%} & 80.313\% \\
AUC Score & \textbf{92.786\%} & 92.596\% & 91.804\% & 91.212\% \\
G-mean & 77.211\% & 80.188\% & \textbf{83.530\%} & 79.041\% \\
\bottomrule
\end{tabular}
\end{table}

\begin{figure}[t]
\vspace*{0.2cm}
  \centering
    \includegraphics[width=0.7\linewidth]{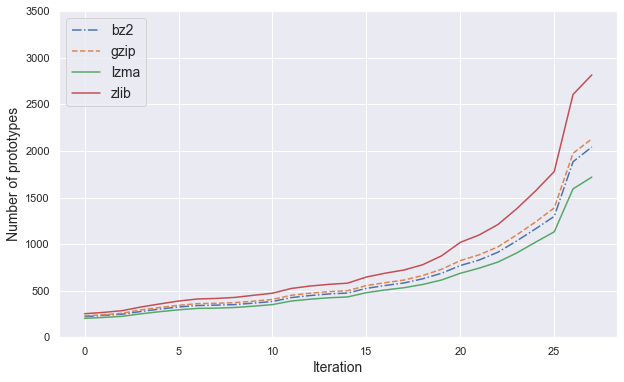}
    \caption{Number of prototypes}
    \label{fig:num_of_prototypes_compression_comparison_1}
\end{figure}

\begin{figure}
  \centering
    \includegraphics[width=0.7\linewidth]{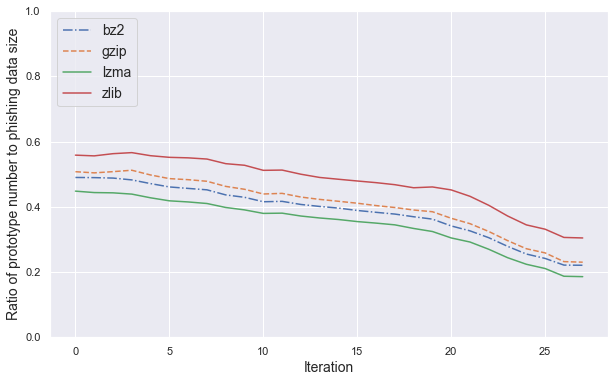}
    \caption{Ratio of prototype number to phishing data size}
    \label{fig:num_of_prototypes_compression_comparison}
\end{figure}

To observe the memory requirements, we also analyzed the number of prototypes extracted at each iteration in an incremental learning experiment in comparison to the number of phishing websites these prototypes represent. Fig.~\ref{fig:num_of_prototypes_compression_comparison_1} shows the number of extracted prototypes, while Fig.~\ref{fig:num_of_prototypes_compression_comparison} depicts the compression ratios at each iteration in the incremental learning experiment (Algorithm~\ref{alg:incremental_learning}). Compression ratio in this context is defined as the ratio between the number of prototypes extracted relative to the total number of phishing websites they represent. As the detection system incrementally learns, the number of prototypes increases as new website samples are learned. However, compression ratio gradually decreases over time, showing that the system was able to extract meaningful phishing prototype representations. As shown in Figure~\ref{fig:num_of_prototypes_compression_comparison_1} and Figure~\ref{fig:num_of_prototypes_compression_comparison}, the use of LZMA algorithms achieved the highest number of extracted prototypes which are representative of the online phishing data. The ability to extract more phishing prototypes consequently led to a better performance over time without the need to retain a large number of phishing prototypes, as shown by the decreasing ratio of prototype number to phishing data size.

\color{black}


% \end{IEEEbiography}


%\newpage



%\vfill


\end{document}